\documentclass[mathpazo]{cicp}
\usepackage[T1,T2A]{fontenc}
\usepackage[utf8]{inputenc}
\usepackage[russian,english]{babel}
\usepackage{geometry}
\usepackage{color}
\usepackage{array}
\usepackage{textcomp}
\usepackage{multirow}
\usepackage{amsbsy}
\usepackage{amstext}
\usepackage{graphicx}
\usepackage{esint}

\begin{document}
%%%%% title : short title may not be used but TITLE is required.
% \title{TITLE}
% \title[short title]{TITLE}
\title{JefiPIC: A 3-D Full Electromagnetic Particle-in-Cell Simulator Based
on Jefimenko's Equations on GPU}

%%%%% author(s) :
% single author:
% \author[name in running head]{AUTHOR\corrauth}
% [name in running head] is NOT OPTIONAL, it is a MUST.
% Use \corrauth to indicate the corresponding author.
% Use \email to provide email address of author.
% \footnote and \thanks are not used in the heading section.
% Another acknowlegments/support of grants, state in Acknowledgments section
% \section*{Acknowledgments}
\author[Jian-Nan Chen et.~al.]{Jian-Nan Chen\affil{1,2},
       Jun-Jie Zhang\affil{1,2}\comma\corrauth, Xue-Ming Li\affil{3}, Hai-Liang Qiao\affil{1,2}, and Yong-Tao Zhao\affil{4}}
\address{\affilnum{1}\ Northwest Institute of Nuclear Technology, Xi'an, 710024,
China. \\
           \affilnum{2}\ State Key Laboratory of Intense Pulsed Radiation Simulation
and Effect, 710024, China. \\
\affilnum{3}\ Institute of Applied Physics and Computational Mathematics,
Beijing, 100094, China. \\
\affilnum{4}\ MOE Key Laboratory for Nonequilibrium Synthesis and Modulation
of Condensed Matter, School of Science, Xi'an Jiaotong University,
Xi'an, 710049, China.}
 \emails{{\tt zjacob@mail.ustc.edu.cn} (J.~Zhang)}

%%%%% Begin Abstract %%%%%%%%%%%
\begin{abstract}
This paper presents a novel 3-D full electromagnetic particle-in-cell
(PIC) code called JefiPIC, which uses Jefimenko's equations as the
electromagnetic (EM) field solver through a full-space integration
method. Leveraging the power of state-of-the-art graphic processing
units (GPUs), we have made the challenging integral task of PIC simulations
achievable. Our proposed code offers several advantages by utilizing
the integral method. Firstly, it offers a natural solution for modeling
non-neutral plasmas without the need for pre-processing such as solving
Poisson's equation. Secondly, it eliminates the requirement for designing
elaborate boundary layers to absorb fields and particles. Thirdly,
it maintains the stability of the plasma simulation regardless of
the time step chosen. Lastly, it does not require strict charge-conservation
particle-to-grid apportionment techniques or electric field divergence
amendment algorithms, which are commonly used in finite-difference
time-domain (FDTD)-based PIC simulations. To validate the accuracy
and advantages of our code, we compared the evolutions of particles
and fields in different plasma systems simulated by three other codes.
Our results demonstrate that the combination of Jefimenko's equations
and the PIC method can produce accurate particle distributions and
EM fields in open-boundary plasma systems. Additionally, our code
is able to accomplish these computations within an acceptable execution
time. This study highlights the effectiveness and efficiency of JefiPIC,
showing its potential for advancing plasma simulations.
\end{abstract}
%%%%% end %%%%%%%%%%%

%%%%% AMS/PACs/Keywords %%%%%%%%%%%
%\pac{}

%%%% maketitle %%%%%
\maketitle
Our code is published at: 

1. https://github.com/Juenjie/JefiPIC 

2. https://codeocean.com/capsule/5129018/tree

%%%% Start %%%%%%
\section{Introduction}
\label{sec1}
\subsection{Background}
The particle-in-cell (PIC) method is a commonly used tool in plasma
physics. It utilizes macro-particles to describe charged particles
in similar phase space states and model the evolution of the particle
distribution and electromagnetic field. This method was initially
proposed by Dawson in the 1960s \cite{Dawson-1962} to study the
Langmuir wave in 1-D electrostatic plasma. Later, Langdon and Birdsall
improved the PIC model by incorporating finite-size particles \cite{Laugdon-1970}
or particle clouds \cite{Birdsall-1969}, which solved the issue
of Coulomb collision between particles. Marder \cite{Marder-1987}
and Villasenor \cite{Villasensor-1992} addressed the electric field
divergence error in the current-driven method. In the last decade, the teams from P. Gibbon and A.J. Christlieb have separately established the integral-method based PIC through solving the vector and scalar potential functions, which can be employed in  electrostatic, magneto-static, and electromagnetic problems and has expanded the study in grid-free plasma simulation \cite{G. Paul-2002}-\cite{A.J.-2016}. Currently, PIC is used
across various fields, including simulating controlled/laser thermonuclear
fusion \cite{Wang-2019},\cite{Arber-2015}, studying nuclear explosions
\cite{Chen-2022}-\cite{Xu-2021} and space physics effects \cite{Lu-2010},
and designing vacuum electronic devices \cite{Chen-2022-1}-\cite{Welch2020}.

The PIC method offers an intuitive representation of charged particles,
making it easier for researchers to analyze simulated phenomena and
data. As a result, numerous commercial software and open-source PIC
codes have emerged for decades, including but not limited to Smilei
\cite{smilei-2018}, PIConGPU \cite{PICONGPU-2010}, Warpx \cite{Warpx-2018},
UNIPIC \cite{Wang-2009}, and EPOCH \cite{Arber-2015},\cite{Ridgers-2014}.

The PIC model comprises two primary components. The first part involves
updating the dynamics of EM fields according to Maxwell's equations,
\begin{eqnarray}
\varepsilon\frac{\partial\mathbf{E}}{\partial t} & = & \nabla\times\frac{\mathbf{B}}{\mu}-\mathbf{J},\label{eq:Maxwell-1}\\
\frac{\partial\mathbf{B}}{\partial t} & = & -\nabla\times\frac{\mathbf{E}}{\mu},\label{eq:Maxwell-2}\\
\nabla\cdot\mathbf{E} & = & \frac{\rho}{\varepsilon},\label{eq:Maxwell-3}\\
\nabla\cdot\mathbf{B} & = & 0,\label{eq:Maxwell-4}
\end{eqnarray}
where $\mathbf{E}$ and $\mathbf{B}$ represent electric and magnetic
fields, $\varepsilon$ and $\mu$ denote the permittivity and permeability
of the medium, $\rho$ and $\mathbf{J}$ are the electric charge density
and electric current density. 

The second part involves updating the positions and velocities of
macro-particles using the Newton-Lorentz force equations of motion,
\begin{eqnarray}
\frac{\mathrm{d}}{\mathrm{d}t}(\gamma m\mathbf{v}) & = & q(\mathbf{E}+\mathbf{v}\times\mathbf{B}),\label{eq:particle_motion-1}\\
\frac{\mathrm{d}}{\mathrm{d}t}\mathbf{r} & = & \mathbf{v},\label{eq:particle_motion-2}
\end{eqnarray}
where $\gamma$, $m$, $q$, $\mathbf{r}$, and $\mathbf{v}$ represent
the relativistic factor, particle's mass, charge, displacement, and
velocity, respectively. 

After ensuring current continuity,
\begin{eqnarray}
\frac{\partial\rho}{\partial t}+\nabla\cdot\mathbf{J} & = & 0,\label{eq:continuity_eq}
\end{eqnarray}
the electric field divergence Eq. (\ref{eq:Maxwell-3}) is implicitly
obtained solely through the first two curl Eqs. (\ref{eq:Maxwell-1})
and (\ref{eq:Maxwell-2}), while Eq. (\ref{eq:Maxwell-4}) holds at
all time. 

\subsection{Motivation}

The finite-difference time-domain (FDTD) method \cite{Yee-1966}
is well-suited for computing curl Eqs. (\ref{eq:Maxwell-1}) and (\ref{eq:Maxwell-2})
due to its simplicity and data sparsity. This makes it more advantageous
than other methods, such as the finite element method (FEM) \cite{Na-2017},\cite{Na-2018}
that necessitates the design of complex basic functions. Thus, the FDTD method
has been employed as the EM solver for several decades. In the FDTD-based
PIC framework, it is crucial that the finite-difference implementation
of the current continuity Eq. (\ref{eq:continuity_eq}) should be
consistent with that of the electric field \textbf{$\mathbf{E}$}
and magnetic field $\mathbf{B}$ \cite{Villasensor-1992}. Furthermore,
this current-driven method computes the electrostatic fields indirectly
by Eq. (\ref{eq:continuity_eq}), rather than directly through the
diverge Eq. (\ref{eq:Maxwell-3}) \cite{Langdon-1992}. To ensure
its accuracy, the current interpolation algorithm is meticulously
designed to uphold the charge-conservation law \cite{Verboncoeur-2001}-\cite{Wang-2016}.

In this paper, we present JefiPIC, an alternative PIC method that
utilizes Jefimenko's equations \cite{Jefimenko-1989},\cite{Griffiths-1999},
the general solution to Maxwell's equations, to compute EM fields
in plasma simulations by the integral form. Jefimenko's equations
are computationally time-consuming, particularly when used in conjunction
with the numerous particles updating in PIC methods. To enhance the
performance of this integral simulations, we employ the GPU-based
package JefiGPU \cite{Zhang-2022}, which directly calculate Jefimenko's
equation on the GPU with a high-dimensional integration package ZMCintegral
\cite{Wu-2020}. This approach allows us to efficiently evolve plasma
systems within an acceptable execution time. Furthermore, we perform
particle motion equations on GPU, making JefiPIC a fully GPU-based
plasma simulator. All variables are consistently defined in GPU memory,
ensuring swift execution of the simulations. Providing the initial
conditions of charged particles (neutral or not) and the background
EM fields, our code is supposed to automatically compute the particles'
space-time distribution as well as the evolution of EM fields.

Compared to the traditional FDTD-based PIC simulation, our proposed
JefiPIC has the following advantages: 

a) The numerical stability of the integral eliminates the need for
the CFL stability condition \cite{Courant-1967}, a criterion in
explicit finite difference methods, to prevent numerical divergence.
This provides more flexibility and robustness in selecting the computational
time step. 

b) Our method does not require the use of complex boundary layers
to cut off the propagation of EM fields \cite{Berenger-1994},\cite{Mur-1981}.
Besides, simply 'killing' particle in the boundary layers will not
result in unintended charge deposition or errors in the electrostatic
components \cite{Pasik-1999},\cite{Jost-1997}. Thus, our method
is well-suited for describing plasma systems with open boundaries.

c) Our method can handle non-neutral plasma without the need for additional
pretreatments, such as solving the Poisson equation \cite{Villasensor-1992},
even for those new to the field, making it easier to treat the initial
conditions.

d) In JefiPIC, EM fields are precisely calculated by the utilization
of both charge and current densities, which strictly adheres to the
charge conservation law at every time step. Thus, instead of using
strict charge conservation algorithms \cite{Esirkepov-2001} or implementing
electric field divergence amendments \cite{Marder-1987},\cite{Langdon-1992},
the simple linear interpolation method can be employed. 

Apart from the above merits, our approach inevitably encounters some
drawbacks in its current version. For example, the large computational
cost, and inconvenience in dealing with the EM boundaries with matters,
etc. We will study these issues to make our code more practical in
the future. 

The paper is structured as follows. Section \ref{sec:Algorithm-and-Implementation}
outlines the detailed design process of JefiPIC, including the Jefimenko's
equations, particle motion, the interpolation method for particles
and fields, and the implementation of JefiPIC on GPU. In Section \ref{sec:Computational-Model-and},
we compare the results obtained from JefiPIC, UNIPIC, EPOCH, and RBG-Maxwell
\cite{Zhang-2020},\cite{Zhang-2022-1} through examining the evolutions
of three different plasma systems. We further analyze the advantages
of JefiPIC compared to other available codes. Finally, in Section
\ref{sec:Conclusion}, we present our conclusions and discuss future
research directions.

\section{Algorithm and Implementation\label{sec:Algorithm-and-Implementation}}

\begin{figure}
\begin{centering}
\includegraphics[scale=0.2]{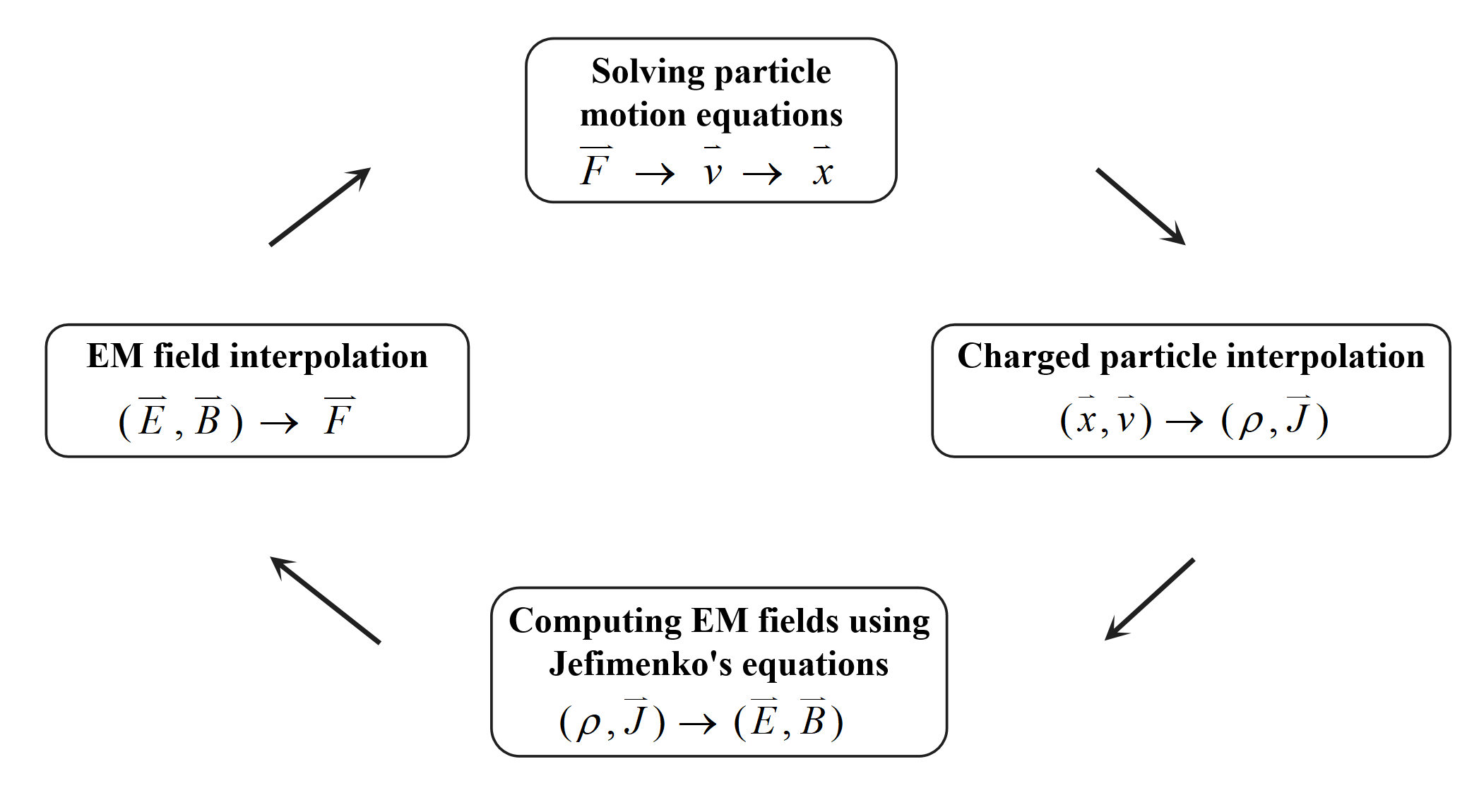}
\par\end{centering}
\caption{Process diagram for JefiPIC simulation. The overall computational
architecture is similar to traditional PIC methods, except for the
use of a different EM field solver and the negligible consideration
of EM field and particle boundary issues.\label{fig:Process-diagram-for}}
\end{figure}

As a PIC method, JefiPIC has a similar architecture to the traditional
EM PIC, shown in Figure \ref{fig:Process-diagram-for}. The method
involves four processes: 1) computing EM fields on each grid, 2) solving
the field interpolation to calculate the forces on the particles,
3) solving the particle interpolation to allocate the charges on the
grids, and 4) updating the particle motion. In this section, we will
introduce the details of JefiPIC in the sequence of Jefimenko's equations,
particle motion, field and particle interpolation and the implementation
on GPU.

\subsection{Jefimenko's equations\label{subsec:Jefimenko's-equations}}

In JefiPIC, the aim is to compute the EM fields over the entire computational
region by the Jefimenko's equations with sources $\mathbf{J}$ and
$\rho$,
\begin{eqnarray}
\mathbf{E}(\mathbf{r},t) & = & \frac{1}{4\pi\varepsilon_{0}}\int\left[\frac{\mathbf{r}-\mathbf{r}^{\prime}}{|\mathbf{r}-\mathbf{r}^{\prime}|^{3}}\rho(\mathbf{r}^{\prime},t_{\mathrm{r}})+\frac{\mathbf{r}-\mathbf{r}^{\prime}}{|\mathbf{r}-\mathbf{r}^{\prime}|^{2}}\frac{1}{c}\frac{\partial\rho(\mathbf{r}^{\prime},t_{\mathrm{r}})}{\partial t}\right.\nonumber \\
 &  & \left.-\frac{\mathbf{r}-\mathbf{r}^{\prime}}{|\mathbf{r}-\mathbf{r}^{\prime}|}\frac{1}{c^{2}}\frac{\partial\mathbf{J}(\mathbf{r}^{\prime},t_{\mathrm{r}})}{\partial t}\right]\mathrm{d}^{3}\mathbf{r}^{\prime},\label{eq:E_Jefi}
\end{eqnarray}
\begin{eqnarray}
\mathbf{B}(\mathbf{r},t)=-\frac{\mu_{0}}{4\pi}\int\left[\frac{\mathbf{r}-\mathbf{r}^{\prime}}{|\mathbf{r}-\mathbf{r}^{\prime}|^{3}}\times\mathbf{J}(\mathbf{r}^{\prime},t_{\mathrm{r}})+\frac{\mathbf{r}-\mathbf{r}^{\prime}}{|\mathbf{r}-\mathbf{r}^{\prime}|^{2}}\times\frac{1}{c}\frac{\partial\mathbf{J}(\mathbf{r}^{\prime},t_{\mathrm{r}})}{\partial t}\right]\mathrm{d}^{3}\mathbf{r}^{\prime},\label{eq:B_Jefi}
\end{eqnarray}
\begin{eqnarray}
t_{\mathrm{r}} & = & t-|\mathbf{r}-\mathbf{r}^{\prime}|/c,\label{eq:retard_time}
\end{eqnarray}
where $(\mathbf{r},t)$ represents the space-time point in the computational
region, while $\mathbf{r}=(x,y,z)$ and $\mathbf{r}^{\prime}=(x^{\prime},y^{\prime},z^{\prime})$
separately refer to the displacements of the EM fields and sources.
For numerical computation, we divide the computational region into
the same structured cuboid grids. All the EM quantities $\mathbf{J}$,
$\rho$, $\mathbf{E}$ and $\mathbf{B}$ are defined at the center
of the grids. To perform integrations numerically, Eqs. (\ref{eq:E_Jefi})
to (\ref{eq:retard_time}) are discretized as,
\begin{eqnarray}
\mathbf{E}(\mathbf{r}_{I,J,K},t^{n}) & = & \frac{\mathrm{d}x^{\prime}\mathrm{d}y^{\prime}\mathrm{d}z^{\prime}}{4\pi\varepsilon_{0}}\sum_{i,j,k}\left[\frac{\mathbf{r}_{I,J,K}-\mathbf{r}_{i,j,k}^{\prime}}{|\mathbf{r}_{I,J,K}-\mathbf{r}_{i,j,k}^{\prime}|^{3}}\rho(\mathbf{r}_{i,j,k}^{\prime},t_{\mathrm{r}})\right.\nonumber \\
 &  & +\frac{\mathbf{r}_{I,J,K}-\mathbf{r}_{i,j,k}^{\prime}}{|\mathbf{r}_{I,J,K}-\mathbf{r}_{i,j,k}^{\prime}|^{2}}\frac{1}{c}\frac{\rho(\mathbf{r}_{i,j,k}^{\prime},t_{\mathrm{r}})-\rho(\mathbf{r}_{i,j,k}^{\prime},t_{\mathrm{r}}-\mathrm{d}t)}{\mathrm{d}t}\nonumber \\
 &  & \left.-\frac{1}{|\mathbf{r}_{I,J,K}-\mathbf{r}_{i,j,k}^{\prime}|}\frac{1}{c^{2}}\frac{\mathbf{J}(\mathbf{r}_{i,j,k}^{\prime},t_{\mathrm{r}})-\mathbf{J}(\mathbf{r}_{i,j,k}^{\prime},t_{\mathrm{r}}-\mathrm{d}t)}{\mathrm{d}t}\right],\label{eq:dis_E_Jefi}
\end{eqnarray}
\begin{eqnarray}
\mathbf{B}(\mathbf{r}_{I,J,K},t^{n}) & = & -\mu_{0}\frac{\mathrm{d}x^{\prime}\mathrm{d}y^{\prime}\mathrm{d}z^{\prime}}{4\pi}\sum_{i,j,k}\left[\frac{\mathbf{r}_{I,J,K}-\mathbf{r}_{i,j,k}^{\prime}}{|\mathbf{r}_{I,J,K}-\mathbf{r}_{i,j,k}^{\prime}|^{3}}\times\mathbf{J}(\mathbf{r}_{i,j,k}^{\prime},t_{\mathrm{r}})\right.\nonumber \\
 &  & \left.+\frac{\mathbf{r}_{I,J,K}-\mathbf{r}_{i,j,k}^{\prime}}{|\mathbf{r}_{I,J,K}-\mathbf{r}_{i,j,k}^{\prime}|^{2}}\times\frac{1}{c}\frac{\mathbf{J}(\mathbf{r}_{i,j,k}^{\prime},t_{\mathrm{r}})-\mathbf{J}(\mathbf{r}_{i,j,k}^{\prime},t_{\mathrm{r}}-\mathrm{d}t)}{\mathrm{d}t}\right],\label{eq:dis_B_Jefi}
\end{eqnarray}
\begin{eqnarray}
t_{r} & = & t^{n}-|\mathbf{r}_{I,J,K}-\mathbf{r}_{i,j,k}^{\prime}|/c,\label{eq:dis_retard_time}
\end{eqnarray}
where the subscripts $I$, $J$, and $K$ in capital letter of displacement denote the
grid index of the unknown EM fields, while the subscripts $i$, $j$,
and $k$ in lowercase letter of displacement denote the grid index of the source. The
superscript n of time refers to the n-th time step, and $\mathrm{d}x^{\prime}$,
$\mathrm{d}y^{\prime}$, and $\mathrm{d}z^{\prime}$ represent the
size of a grid. 

\subsection{Particle motion\label{subsec:Particle-motion}}

The Newton\textendash Lorentz force Eqs. (\ref{eq:particle_motion-1})
to (\ref{eq:particle_motion-2}) are discretized by the central leap-frog
difference method as,
\begin{eqnarray}
\frac{\gamma m(\mathbf{v}^{n+1/2}-\mathbf{v}^{n-1/2})}{\mathrm{d}t} & = & q\left(\mathbf{E}^{n}+\frac{\mathbf{v}^{n+1/2}+\mathbf{v}^{n-1/2}}{2}\times\mathbf{B}^{n}\right),\label{eq:dis_particle_motion-1}\\
\frac{\mathbf{r}^{n+1}-\mathbf{r}^{n}}{\mathrm{d}t} & = & \mathbf{v}^{n+1/2},\label{eq:dis_particle_motion-2}
\end{eqnarray}
This advancement of velocity involves an implicit equation. To figure
out this issue, the Boris particle pusher is employed to separate
Eq. (15) into three steps as,
\begin{eqnarray}
\mathbf{v}^{-} & = & \mathbf{v}^{n-1/2}+\frac{q\mathbf{E}^{n}}{m}\frac{\mathrm{d}t}{2},\label{eq:v_update_1}\\
\frac{(\mathbf{v}^{+}+\mathbf{v}^{-})}{\mathrm{d}t} & = & \frac{q}{2\gamma m}(\mathbf{v}^{+}+\mathbf{v}^{-})\times\mathbf{B}^{n},\label{eq:v_update_2}\\
\mathbf{v}^{n+1/2} & = & \mathbf{v}^{+}+\frac{q\mathbf{E}^{n}}{m}\frac{\mathrm{d}t}{2},\label{eq:v_update_3}
\end{eqnarray}
where $\mathbf{v}^{-}$ and $\mathbf{v}^{+}$ denote the temporary
variables to update the velocity. 

\subsection{Charge, current and field interpolation method\label{subsec:Charge,-current-and}}

Based on the discretization of Jefimenko's equations, the EM fields,
current and charge densities all need to be allocated at the center
of the grids, which is different from that in the FDTD-based method. The traditional
PIC methods usually adopt two kinds of interpolation methods. One
approach is the \textit{charge-conservation} method \cite{Buneman-1968},
which generate high-frequency numerical noises in the solenoidal part of current
density \cite{Marder-1987},\cite{Langdon-1992},\cite{Goplin-1983}.
The other method is the sample \textit{linear interpolation} method,
which generates less noise \cite{Marder-1987}, but violates the
charge conservation law, leading to inaccuracies in electric field.\textcolor{magenta}{{}
}To compensate for this loss in accuracy, electric field correction
algorithms must be implemented.

Because the integral-based PIC method calculates the fields directly
from the provided current and charge density, JefiPIC avoids the issue
of charge conservation. As a result, we can use the first-order linear
interpolation method for charge and field allocation in JefiPIC, combining
the advantages of both acceptable numerical noise and computational
time.

\begin{figure}
\begin{centering}
\includegraphics[scale=0.3]{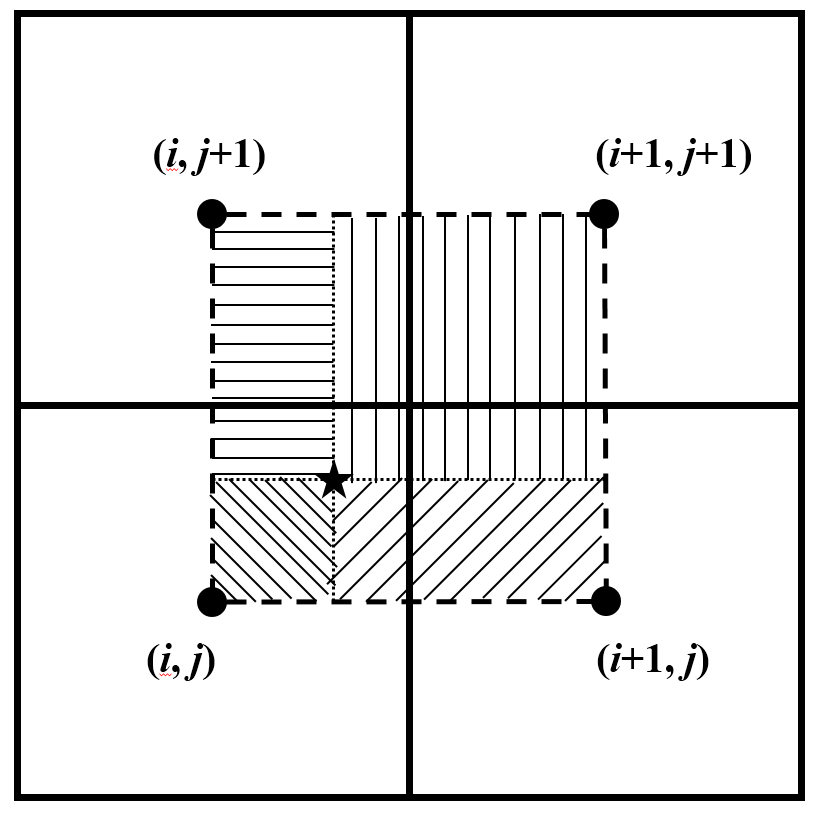}
\par\end{centering}
\caption{Area-weighted interpolation. The particle is indicated by a 'star'
and its charge is interpolated to the four neighboring grids according
to the ratio of the corresponding shaded area to a single grid area.\label{fig:Area-weighted-interpolation.-The}}
\end{figure}

We firstly exhibit how particle interpolation works. In a 3-D model,
linear interpolation is achieved through volume-weighted interpolation.
For simplicity, we use a 2-D model with the area-weighted interpolation
method, as shown in Figure \ref{fig:Area-weighted-interpolation.-The}.
Assuming that there is one particle in the computational region, labeled
as a 'star'. The particle's charge is distributed to the nearest four
center nodes indicated by the circles with indices $(i,j)$, $(i+1,j)$,
$(i,j+1)$ and$(i+1,j+1)$. The area-weighted interpolation method
specifies that the allocated charge on each node is proportional to
the weight, which is the ratio of the area of its opposite rectangle
to that of a single grid. Accordingly, the weight can be computed
by the corresponding volume in 3-D models. The zero-, first- and 
second-order spline interpolations are as follows,
\begin{equation}
S^{(0)}(x)=1\ \text{\ for}\ \ 0\leq|x|\leq\Delta
\end{equation}

\begin{equation}
S^{(1)}(x)=\frac{\Delta-x}{\Delta}\ \text{\ for}\ \ 0\leq|x|\leq\Delta
\end{equation}

\begin{equation}
S^{(2)}(x)=\begin{cases}
\begin{array}{c}
\frac{1}{\Delta^{2}}\left(-x^{2}+\frac{3}{4}\Delta^{2}\right)\\
\frac{1}{8\Delta^{2}}\left(2x-3\Delta\right)^{2}
\end{array} & \begin{array}{c}
\text{for}\ \ 0\leq|x|\leq\Delta\\
\text{\ \ \ \ for}\ \ \frac{1}{2}\Delta\leq|x|\leq\frac{3}{2}\Delta
\end{array}\end{cases}
\end{equation}

Since all the physical variables are defined at the center nodes of
the grids, we firstly need to find the indices of the grids which
the particles belong to,
\begin{eqnarray}
i=\left[\frac{x-0.5\cdot\mathrm{d}x^{\prime}}{\mathrm{d}x^{\prime}}\right], j=\left[\frac{y-0.5\cdot\mathrm{d}y^{\prime}}{\mathrm{d}y^{\prime}}\right], k=\left[\frac{z-0.5\cdot\mathrm{d}z^{\prime}}{\mathrm{d}z^{\prime}}\right],\label{eq:indices}
\end{eqnarray}
where the square bracket means rounding down operation, and $x,$
$y$, and $z$ are the displacements in the three axes. Following
the volume-weighted interpolation method, a particle located at one
grid $(i,j,k)$ should interpolate its charge to the adjacent grid
nodes as,
\[
Q_{i,j,k}=(1-\alpha)(1-\beta)(1-\chi)q,
\]

\[
Q_{i+1,j,k}=\alpha(1-\beta)(1-\chi)q,
\]

\[
Q_{i,j+1,k}=(1-\alpha)\beta(1-\chi)q,
\]

\[
Q_{i,j,k+1}=(1-\alpha)(1-\beta)\chi q,
\]

\[
Q_{i+1,j+1,k}=\alpha\beta(1-\chi)q,
\]

\[
Q_{i+1,j,k+1}=\alpha(1-\beta)\chi q,
\]

\[
Q_{i,j+1,k+1}=(1-\alpha)\beta\chi q,
\]

\begin{equation}
Q_{i+1,j+1,k+1}=\alpha\beta\chi q,\label{eq:interpolation_method}
\end{equation}
where
\begin{eqnarray}
\alpha=\frac{x-(i+0.5)\cdot\mathrm{d}x^{\prime}}{\mathrm{d}x^{\prime}},  \beta=\frac{y-(j+0.5)\cdot\mathrm{d}y^{\prime}}{\mathrm{d}y^{\prime}},  \chi=\frac{z-(k+0.5)\cdot\mathrm{d}z^{\prime}}{\mathrm{d}z^{\prime}},\label{eq:parameters}
\end{eqnarray}
and $q$ is the charge of the moving particle. We can acquire the
charge and current density at node $(i,j,k)$ by,
\begin{eqnarray}
\rho_{i,j,k} & = & \frac{Q_{i,j,k}}{\mathrm{d}x^{\prime}\cdot\mathrm{d}y^{\prime}\cdot\mathrm{d}z^{\prime}},\label{eq:def_rho}\\
\left(J_{i,j,k,x},J_{i,j,k,y},J_{i,j,k,z}\right) & = & \rho_{i,j,k}\cdot(v_{x},v_{y},v_{z}),\label{eq:def_J}
\end{eqnarray}
where $v_{x}$, $v_{y}$ and $v_{z}$ are the velocities of the particle.
The charge and current densities at the corresponding nodes can be
obtained according to Eqs. (\ref{eq:interpolation_method}) to (\ref{eq:def_J}).

\begin{figure}
\begin{centering}
\includegraphics[scale=0.2]{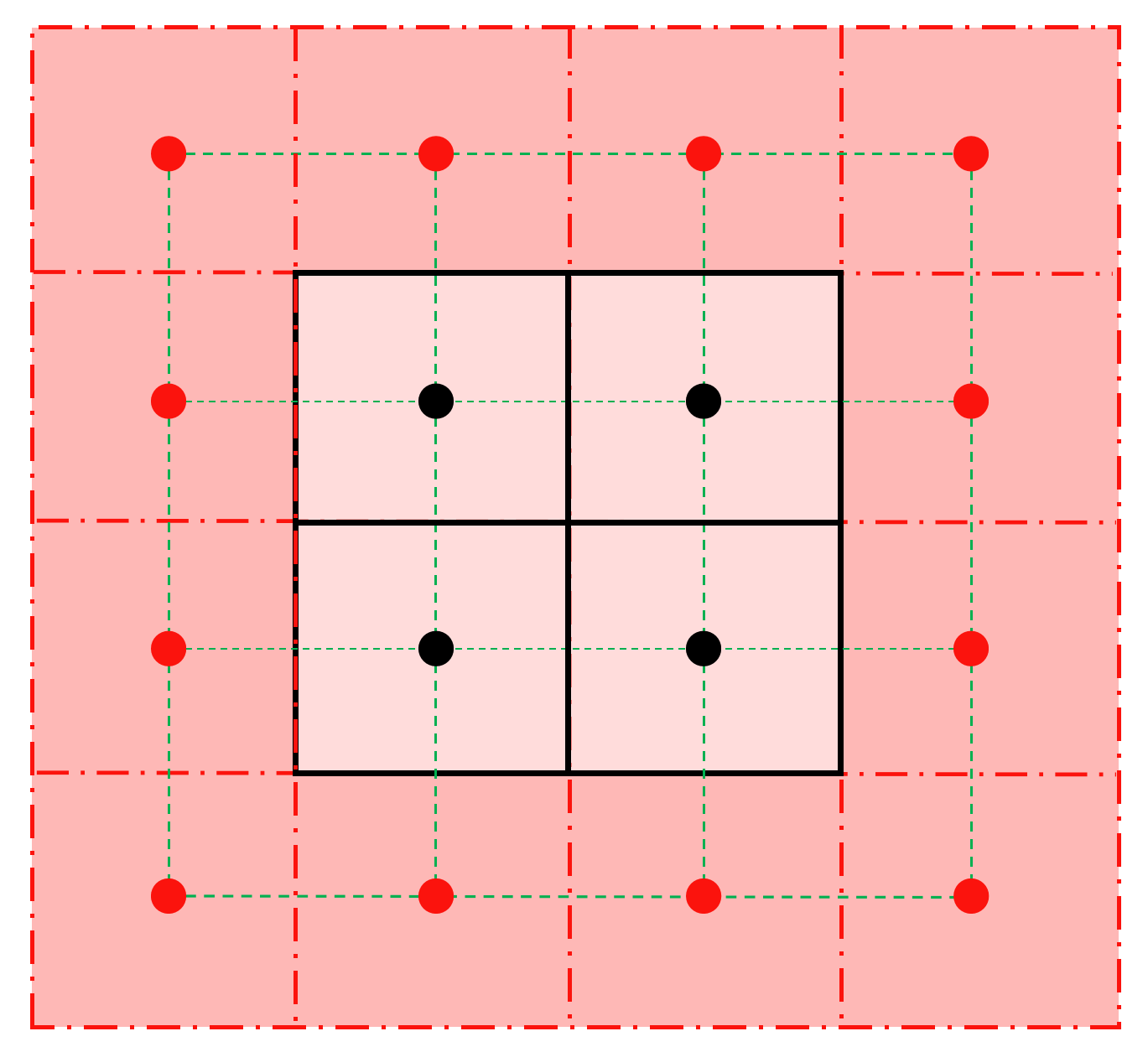}
\par\end{centering}
\caption{Virtual grids in the charge allocation. The black grids and nodes
represent the real computational region, while the red dashed grids
and nodes are in the virtual one. The linked green dashed grids are
used for regular 2-D charge interpolation as in Figure 2. During this
process, only the charge densities on the black nodes will be computed
and saved.\label{fig:Virtual-grids-in}}
\end{figure}

When a particle is located closer to a computational region boundary
than half a grid length, it is important to note that the adjacent
nodes cannot constitute a full grid. To address this issue, we utilize
virtual grids (red grids) to surround the computational region (black
grids), as illustrated in Figure \ref{fig:Virtual-grids-in}. We can
then implement the same charge interpolation process as displayed
in Figure \ref{fig:Area-weighted-interpolation.-The}, while excluding
the computation of charge and current densities on the virtual nodes.

In this part, we will show the field interpolation. Once the EM fields
are established on all observation grids, it is required to compute
the forces on the particles using the field interpolation method.
The field interpolation process works in reverse to that of the charge
interpolation method, whereby the weighted EM fields from the adjacent
eight grids are summed to calculate the fields acting on a certain
particle. The weight of electric field from one grid on the particle
is equal to the weight of the particles' charge allocated to that
corresponding grid. Thus, the electric field on a particle located
in grid $(i,j,k)$ can be evaluated as,
\begin{eqnarray}
\mathbf{E}_{i,j,k}^{q} & = & (1-\alpha)(1-\beta)(1-\chi)\mathbf{E}_{i,j,k}+\alpha(1-\beta)(1-\chi)\mathbf{E}_{i+1,j,k}\nonumber \\
 &  & +(1-\alpha)\beta(1-\chi)\mathbf{E}_{i,j+1,k}+(1-\alpha)(1-\beta)\chi\mathbf{E}_{i,j,k+1}\nonumber \\
 &  & +\alpha\beta(1-\chi)\mathbf{E}_{i+1,j+1,k}+\alpha(1-\beta)\chi\mathbf{E}_{i+1,j,k+1}\nonumber \\
 &  & +(1-\alpha)\beta\chi\mathbf{E}_{i,j+1,k+1}+\alpha\beta\chi\mathbf{E}_{i+1,j+1,k+1}.\label{eq:E_interpolation}
\end{eqnarray}
The dimensionless parameters in Eq. (\ref{eq:E_interpolation}) correspond
to those in Eq. (\ref{eq:parameters}), thus allowing for the computation
of the magnetic field $\mathbf{B}$ on the particle through the same
method. It is worth noting that the number of adjacent nodes involved
in driving the particle follows the same rule as particle interpolation,
and the EM fields outside the computational region are not calculated.
Thus, once the particles pass through the computational boundary,
the forces acting upon them are no longer considered, and they do
not return to the computational region. As we do not solve the charge
conservation equation, we can exclude their contribution to the source
without resulting in undesired charge deposition.

\subsection{GPU Implementation}

According to sections \ref{subsec:Jefimenko's-equations} to \ref{subsec:Charge,-current-and},
JefiPIC needs three major CUDA kernels, each taking over of the Jefimenko's
equation, particle motion, and the charge and field interpolation,
respectively.

The first CUDA kernel is designed to perform the summations presented
in Eqs. (\ref{eq:dis_E_Jefi}) to (\ref{eq:dis_retard_time}) on GPU.
Since the EM fields of a grid are produced by all $\mathbf{J}$ and
$\rho$ in the computational region with retarded time $t_{r}$, we
need to figure out the quantity of sources that should be saved. Here
is our solution. Firstly, we calculate the maximum length ($L_{\text{max}}$)
across the computational region, while the value of $L_{\text{max}}$
is typically determined by the diagonal length of the computational
region. We then divide $L_{\text{max}}$ by $c\mathrm{d}t$ to obtain
$N_{\mathrm{d}t}=\text{Celing}(L_{\text{max}}/(c\mathrm{d}t))$, which
represents the maximum number of time steps an EM field travelling
across the entire computational region. Consequently, the sources
$\mathbf{J}$ and $\rho$ for at least the most recent $N_{\mathrm{d}t}$
time steps need to be stored to guarantee the EM field computation.
For instance, at time $t_{\mathrm{A}}$, we need to save $\mathbf{J}$
and $\rho$ at all grids (the number of grids is M) for $N_{\mathrm{d}t}$
time steps, i.e.,

\noindent $t_{\mathrm{A}}-\mathrm{d}t:[\rho(r_{1},t_{\mathrm{A}}-\mathrm{d}t),J(r_{1},t_{\mathrm{A}}-\mathrm{d}t)],\ldots,[\rho(r_{\mathrm{M}},t_{\mathrm{A}}-\mathrm{d}t),J(r_{\mathrm{M}},t_{\mathrm{A}}-\mathrm{d}t)],$

\noindent $t_{\mathrm{A}}-2\cdot\mathrm{d}t\text{:}[\rho(r_{1},t_{\mathrm{A}}-2\cdot\mathrm{d}t),J(r_{1},t_{\mathrm{A}}-2\cdot\mathrm{d}t)],\ldots,[\rho(r_{\mathrm{M}},t_{\mathrm{A}}-2\cdot\mathrm{d}t),J(r_{\mathrm{M}},t_{\mathrm{A}}-2\cdot\mathrm{d}t)],$

\noindent $\cdots\cdots$

\noindent $t\mathrm{_{A}}-N_{\mathrm{d}t}\cdot\mathrm{d}t:[\rho(r_{1},t\mathrm{_{A}}-N_{\mathrm{d}t}\cdot\mathrm{d}t),J(r_{1},t\mathrm{_{A}}-N_{\mathrm{d}t}\cdot\mathrm{d}t)],\ldots,[\rho(r_{\mathrm{M}},t\mathrm{_{A}}-N_{\mathrm{d}t}\cdot\mathrm{d}t),J(r_{\mathrm{M}},t\mathrm{_{A}}-N_{\mathrm{d}t}\cdot\mathrm{d}t)].$

For time steps $N$ that are ahead of the initial time $t_{0}(t\mathrm{_{A}}-N\cdot\mathrm{d}t<t_{0})$,
$\mathbf{J}$ and $\rho$ are simply set to zero. Here $\mathbf{J}$
and $\rho$ are cached in the global GPU memory. Obviously, this procedure
requires a large amount of memory space on GPU. In practice, we limit
$N_{\mathrm{d}t}$ to 10000, i.e., $N_{\mathrm{d}t}=\text{min}(N_{\mathrm{d}t},10000)$,
making the data storage achievable for most GPU apparatus.

Though we use a truncation level of 10000 to account for the time history, $N_{\mathrm{d}t}$ is frequently much smaller than it. Hence, with this approach, almost all sources can be stored in the GPU to compute the fields, thus ensuring the conservation properties in most scenarios. The cases where the value of $L_{\text{max}}$ by $c\mathrm\cdot{d}t$ exceeds 10000 generally correspond to unrealistic scenarios with very large computational regions or very small time steps. For the former, the influence of distant fields on the observation point would be minimal. For the latter, it negates the advantage of JefiPIC, which is the ability to use larger time steps. Even if the truncation of integral time may introduce minor charge conservation error, this error will not accumulate owing to the integral method.

\begin{figure}
\begin{centering}
\includegraphics[scale=0.2]{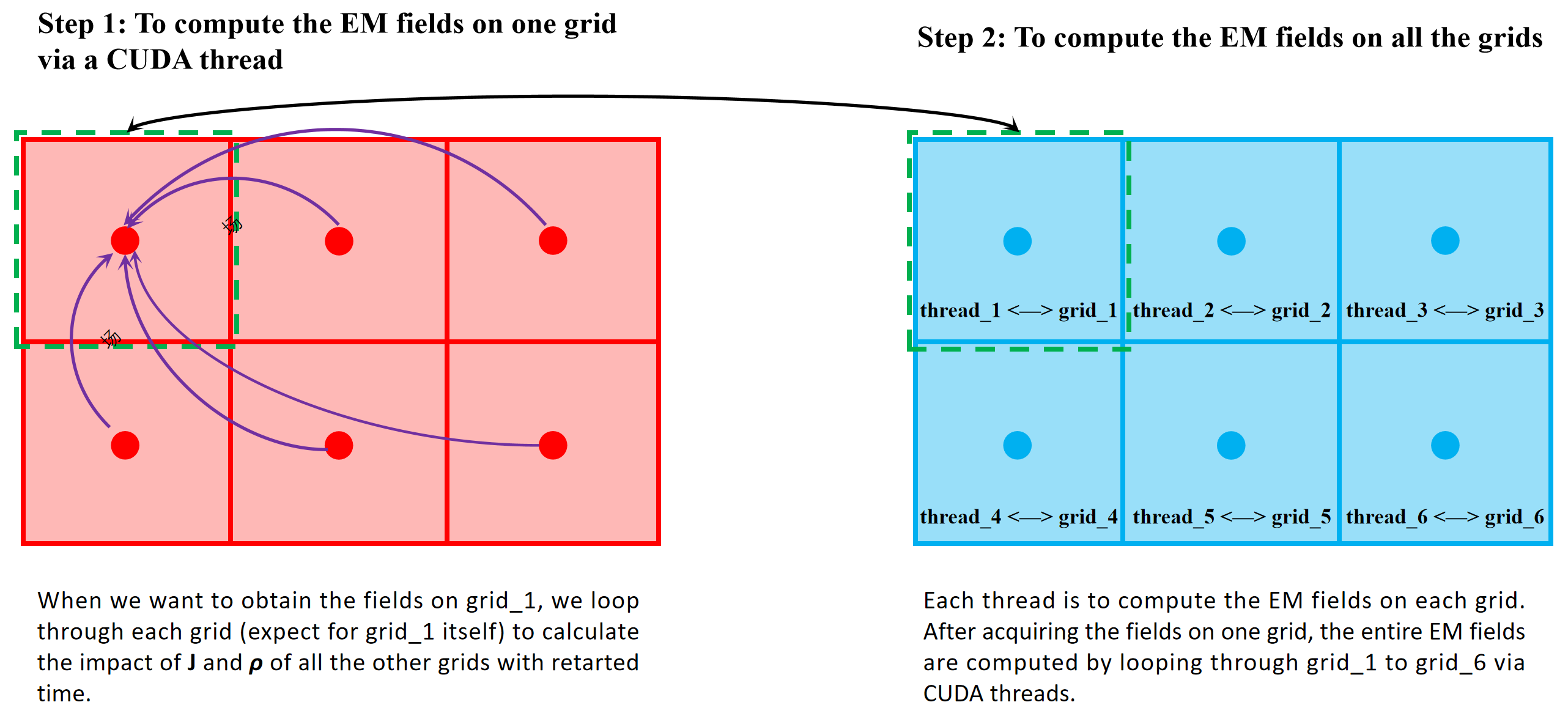}
\par\end{centering}
\caption{Process of the implementation of solving Jefimenko's equations on
GPU. Supposing that the computational region contains 6 grids, and
the process involves two steps. In the first step, EM fields on one
grid (for example $grid\text{\_}i$) are computed by looping through
the other grids with $\mathbf{J}$ and $\rho$. In the second step,
the entire EM fields are computed by looping through the rest grids
repeating the operation in step 1.\label{fig:Process-of-the}}
\end{figure}

In the implementation of the integral, each thread in the CUDA kernel
calculates the EM fields corresponding to a certain grid. Thus, we
loop through each grid in the computational region. For convenience,
we denote the $i-th$ CUDA thread as $thread\text{\_}i$, which calculates
the EM fields on the $i-th$ grid with index $grid\text{\_}i$. To
evaluate the EM fields of $grid\text{\text{\_}}i$, we loop through
all the grids to compute the Jefimenko's equation through the local
sources $\boldsymbol{\mathbf{J}}$ and $\rho$ with retarded time
tr (except for $grid\text{\_}i$ itself to prevent numerical divergence)
and sum all the EM fields. This procedure is depicted in Figure \ref{fig:Process-of-the}.

The integrative form with a record of $N_{\mathrm{d}t}$ moments is beneficial for managing low-frequency electromagnetic effects. This approach sets our method apart from the quasi-static Darwin model, which presumes instantaneous transmission of information, thereby eliminating any time delay. Consequently, our proposed solver transcends the constraints of low-frequency situations as an electromagnetic solver. Nonetheless, to thoroughly account for the characteristics of electromagnetic wave propagation, it is still necessary to track the historical data while using integration methods. Furthermore, our method necessitates a mesh to carry out the integrals required to determine the charge densities defined on the mesh. This approach differs from the mesh-free method \cite{G. Paul-2017}, where the electromagnetic fields, impacting a specific macro-particle, are ascertained by a weighted summation of other particles rather than the charge densities on the grid. Moreover, the integral equation employed in our study is entirely based on the retarded potential approach, therefore is different from the integral method of convolution operators, a stabilized version of the finite difference method. For instance, the work carried out by Wolf and his colleagues \cite{A.J.-2016} presents a new integral method that eradicates the CFL restriction and does not necessitate the tracking of historical data.

The second CUDA kernel is to achieve the particle motion exhibited
by Eqs. (\ref{eq:dis_particle_motion-1}) to (\ref{eq:v_update_2}).
Here, the particle's displacements for full time steps and velocities
for half time steps, i.e., $r^{n}$, $v^{n-1/2}$, are saved in the
global GPU memory. The information update of each particle is also
handled by an individual thread in the CUDA kernel. 
Hence, the maximum particle numbers allowed in our code is restricted by the maximum size of each dimension of a grid of thread blocks (one can simply understand it as the maximum number of threads). For one NVIDIA A100 card, this number is about $10^{9}$, meaning that we cannot calculate more than $10^{9}$ particles on one GPU card.

Since $\mathbf{J}$, $\rho$, $\mathbf{E}$ and $\mathbf{B}$ are
defined at the grid center node, we need to build the third CUDA kernel
to calculate the forces on the particles and the charge and current
densities on all the grid centers. To obtain $\mathbf{E}$ and $\mathbf{B}$,
we first iterate through all the particles in the computational region
to compute $\mathbf{J}$ and $\rho$ for each grid. We then use the
known $\mathbf{J}$ and $\rho$ with a retarded time to calculate
$\mathbf{E}$ and $\mathbf{B}$ by iterating through all the grids.
Here, $\mathbf{J}$, $\rho$, $\mathbf{E}$ and $\mathbf{B}$ are
cached in the global GPU memory and all intermediate variables like
$\alpha$ and $\beta$ are set temporary local variables.

For plasma systems at different length scales, the numerical value
of physical quantities can vary significantly. Given that the numerical
values in JefiPIC must lie within the range of machine precision (float64),
we must convert these quantities from SI unit system to a new unit
system. In this paper, we use the Flexible Unit (FU) system (a brief
introduction of FU is presented in the appendix), with which our code
can simulate plasma systems at different length scales on GPU. By
assigning proper values for constants such as vacuum permittivity
$\varepsilon_{0}$, reduced Planck constant $\hbar$, speed of light
$c$, and $\lambda$ (which relates the energy in SI and FU), we can
limit most of the numerical values to machine precision. \textbf{Note
that the use of FU does not change the physics of the plasma system.}

\section{Computational Model and Results\label{sec:Computational-Model-and}}

To exhibit the comprehensive capabilities of JefiPIC, we performed
a comparative study of three plasma models against three other 3-D
codes. We simulated three non-neutral electron plasma models, including
a) electrons emitted from a point to verify the correctness of the
particle and field cut-off boundaries, b) electrons emitted with zero
initial velocity to demonstrate the ease to handle non-neutral plasma,
and c) electrons emitted with random initial velocity, showcasing
the natural charge conservation feature. The alternative codes we
used include UNIPIC, a mature PIC code, EPOCH, an open-source PIC
code, and RGB-Maxwell, a plasma simulator. UNIPIC and EPOCH are traditional
FDTD-based PIC codes, while RGB-Maxwell solves the Boltzmann equations.
Here, we list some presets of these codes.

\textbf{Particle boundary condition} \textemdash{} The three PIC codes
all use the cut-off boundary to 'kill' the particles that leave the
computational region, and RGB-Maxwell directly estimate the particle
distribution function outside the computational region.

\textbf{Field boundary condition} \textemdash{} Codes based on difference
method utilize convolutional perfect matched layers (CPML) as the
field boundary \cite{CPML-1}-\cite{CPML-2}, while those based on integral method use a simple
cut-off boundary. 

\textbf{Space grid and time step} \textemdash{} All four codes divide
the computational region with the same spatial grid size of $\mathrm{d}x=\mathrm{d}y=\mathrm{d}z=10^{-5}$
m. The total computational time is set to 1 ns. UNIPIC and EPOCH use
the difference methods, thus requiring the time step $\mathrm{d}t$
to at least satisfy the CFL condition,
\begin{eqnarray}
c\cdot\mathrm{d}t & \leq & \frac{1}{\sqrt{\frac{1}{(\mathrm{d}x)^{2}}+\frac{1}{(\mathrm{d}y)^{2}}+\frac{1}{(\mathrm{d}z)^{2}}}},\label{eq:CFL_condition}
\end{eqnarray}
limiting $\mathrm{d}t$ to no more than around $1.92\times10^{-14}$
s. JefiPIC and RGB-Maxwell utilize Jefimenko's equations instead,
which are not limited by numerical stability. Therefore, we set the
time step as $\mathrm{d}t=10^{-13}$ s, which is five times larger than
that used in difference methods.

\textbf{Particle} \textemdash{} To efficiently achieve the computation, we used macro-particle to represent electrons in the similar phase space. The ways to handle macro-particles are different in different codes. JefiPIC implements individual tracking for all macro-particles using CUDA kernels on an NVIDIA A100 GPU card, which has a maximum limit of around $10^{9}$ CUDA threads. Therefore, we are limited to running simulations with a maximum of $10^{9}$ particles at once. To reduce the numerical and statistic noise, we use $~10^{7}$ particles in the JefiPIC simulation. EPOCH can automatically merge and split the weights of the particles ensuring an approximate count of 100 particles in each grid to keep the statistical error during the simulation. In UNIPIC, a fixed particle weight leading to at least $~10^{5}$ particles is set to balance numerical noise and execution time before the simulation. On the other hand, RGB-Maxwell use a distribution function on a six-dimensional phase space instead of the concept of 'particle'. The initial velocity is represented in the units of m/s in our simulation, and the energy corresponding to the velocities $v_{0}$ used in the following experiments equals to 10 eV.

\emph{Note that: Though JefiPIC is able to handle 3D projects, we execute 2D-like or 1D problems in our following simulations for the better comprehension and explanation of JefiPIC.}

\subsection{Perfect Absorbing Layer}

The boundary of PIC contains two components: the field boundary and particle boundary. In the difference-based PIC, the accuracy of the EM fields is highly sensitive to the choice of field boundary layers, and simple methods for 'kill' particles in the boundary layer can introduce undesirable electrostatic field errors. JefiPIC, on the other hand, effectively handles the absorption of fields and particles at the computational boundary with a simple cut-off boundary, aided by the integral operation. A comparative simulation has been conducted between JefiPIC and RGB-Maxwell.

\begin{figure}
\begin{centering}
\includegraphics[scale=0.9]{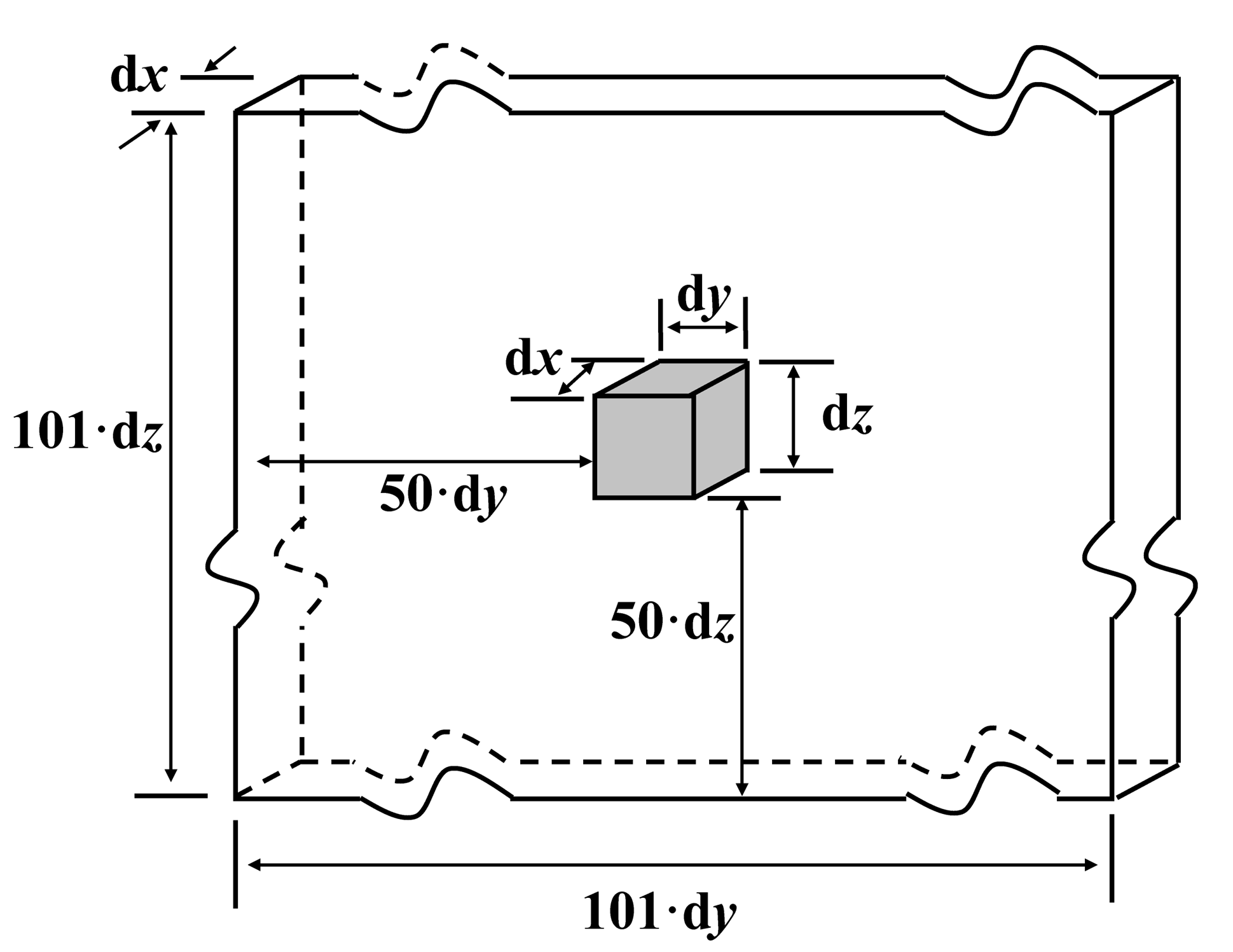}
\par\end{centering}
\caption{Initial particle distribution. The computational region was of size
$1\cdot\mathrm{d}x\times101\cdot\mathrm{d}y\times101\cdot\mathrm{d}z$
in JefiPIC and RGB-Maxwell. The particles are placed in the shaded
region uniformly in the cuboid volume of $1\cdot\mathrm{d}x\times1\cdot\mathrm{d}y\times1\cdot\mathrm{d}z$
ranging from $50\cdot\mathrm{d}y$ to $51\cdot\mathrm{d}y$ in the
$y-\mathrm{axis}$ and $50\cdot\mathrm{d}z$ to $51\cdot\mathrm{d}z$
in the $z-\mathrm{axis}$. The velocities of the particles obey the
Gaussian distribution in the $yoz$ plane.\label{fig:Initial-particle-distribution.}}
\end{figure}

In the two integral PIC code, the total charge of electrons is of $-5\times10^{-14}$ C. They are emitted in all directions in the $y$o$z$ plane from a central point $(i=1,j=50,k=50)$, initialized with a Gaussian distribution of velocities with an average velocity of $v_{0}$ and a standard deviation of 0.2$v_{0}$. This model is simulated by JefiPIC and compared to RGB-Maxwell with the computational region size as $nx\cdot ny\cdot nz = 1\times101\times101$ shown in Figure \ref{fig:Initial-particle-distribution.}.

\begin{figure}
\begin{centering}
\includegraphics{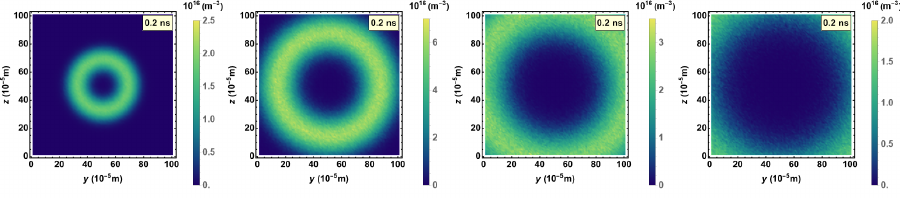}
\par\end{centering}
\begin{centering}
\includegraphics{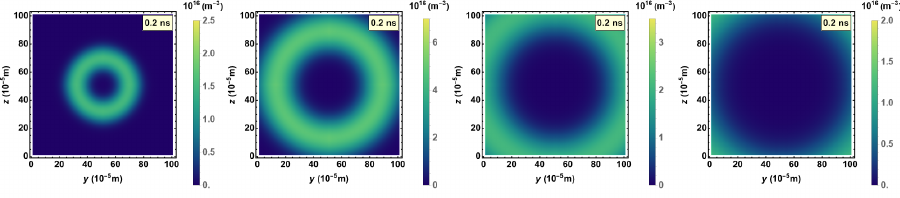}
\par\end{centering}
\caption{Particle distributions over time. Panel (a) shows the distribution
at 0.2 ns, followed by (b) at 0.4 ns, (c) at 0.6 ns, and (d) at 0.8
ns. The two rows represent the particle distributions obtained from
JefiPIC and RGB-Maxwell, respectively.\label{fig:Particle-distributions-over}}
\end{figure}

Figure \ref{fig:Particle-distributions-over} depicts the particle
distribution of the two models at different time instances. The top
row displays the results from the particle model, while the bottom
row shows those from the distribution function. The figure demonstrates
that the two particle distributions are nearly identical, with most
of the particles forming a prominent light ring that spreads out gradually
over time. This outcome confirms the general results obtained from
JefiPIC. Moreover, since the integral method naturally filters out
particles that are outside the computational domain, no charge deposition
occurs when the particles pass through the computational boundary,
and any electromagnetic force on those escape particles would not
be taken into account. Consequently, the motion of the particles is
not hindered, and the ring-like particle distribution is sustained.

\begin{figure}
\begin{centering}
\includegraphics[scale=0.4]{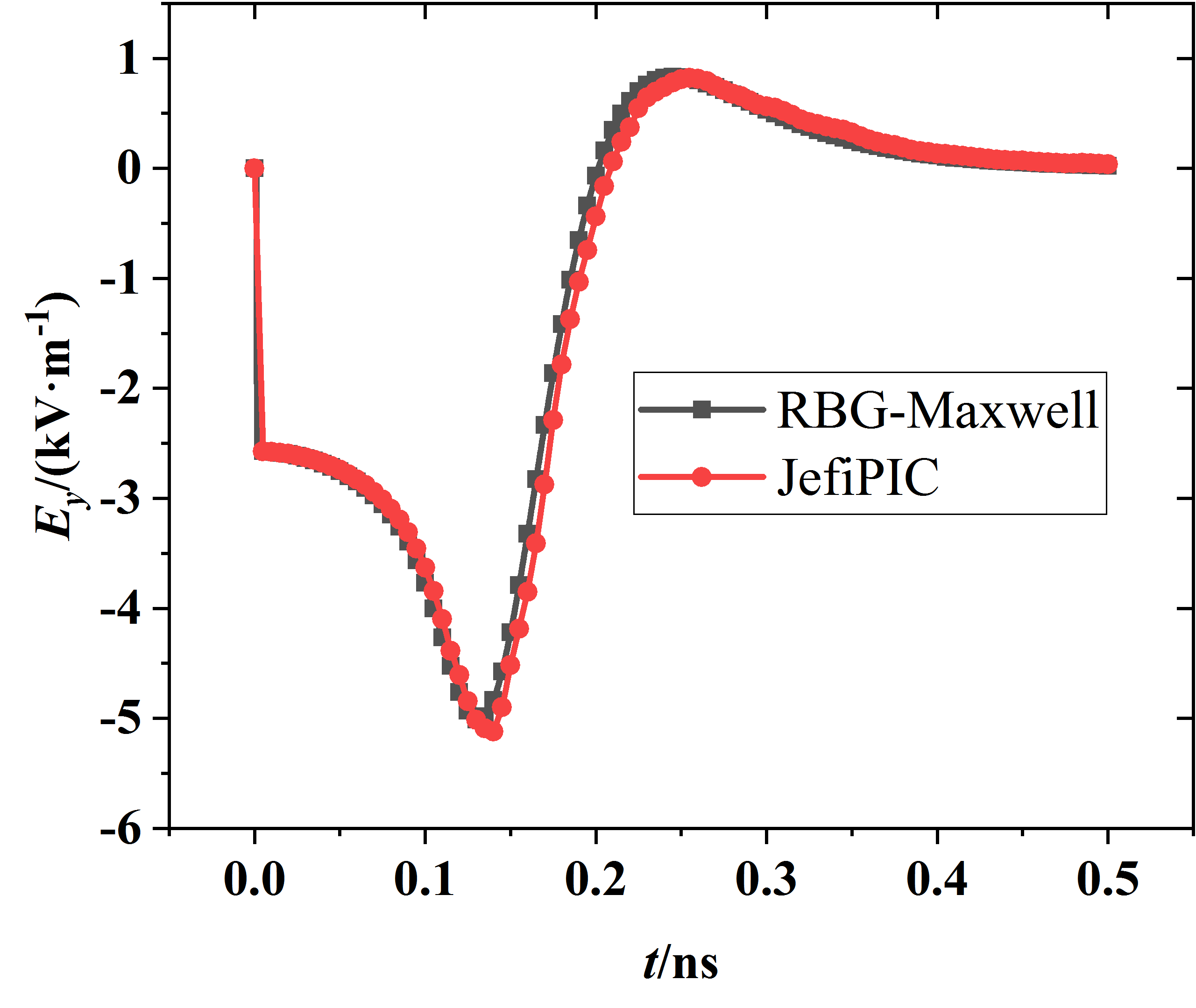}
\par\end{centering}
\caption{Comparison of y-direction electric field. The $y-\mathrm{direction}$
electric fields at point $(i=1,j=75,k=75)$ derived from JefiPIC and
RGB-Maxwell are compared. The black square and red circle represent
the results from RGB-Maxwell and JefiPIC.\label{fig:Comparison-of-y-direction}}
\end{figure}

In Figure \ref{fig:Comparison-of-y-direction}, the electric fields
of the two models are plotted at grid $(i=1,j=75,k=75)$, which is
$25\cdot\mathrm{d}x$ and $25\cdot\mathrm{d}y$ distant from the particle
emission point. As particles spread towards the diagnostic point,
the electric field initially becomes negative and then transitions
to positive before gradually decreasing towards zero as most of the
particles pass and move away from the diagnostic point. The electric
fields in both models display similar waveforms that are smooth with
no observable reflections, suggesting that the electric field does
not interfere with the particle distribution.

Through Figure \ref{fig:Particle-distributions-over} and Figure \ref{fig:Comparison-of-y-direction},
we can observe that despite utilizing the first-order charge and field
interpolation method, JefiPIC delivers results as accurate as those
obtained from the second-order accurate plasma simulator RGB-Maxwell.

As a result, the integral-based PIC method JefiPIC is capable of naturally
cutting off fields and particles at the computational boundary without
reflecting the fields or causing charge deposition errors. This makes
JefiPIC a more effective approach compared to the difference-based
PIC methods that rely on boundary conditions. Thus, JefiPIC is particularly
well-suited for modeling open-boundary problems, such as space plasma
\cite{Howes-2018}, quark-gluon plasma \cite{Zhang-2022-1}, or
high-altitude nuclear explosions \cite{Peng-2021}-\cite{Chen-2019}.

\subsection{Electrostatic effect}
\subsubsection{Non-neutral plasma}

\begin{figure}
\begin{centering}
\includegraphics[scale=0.06]{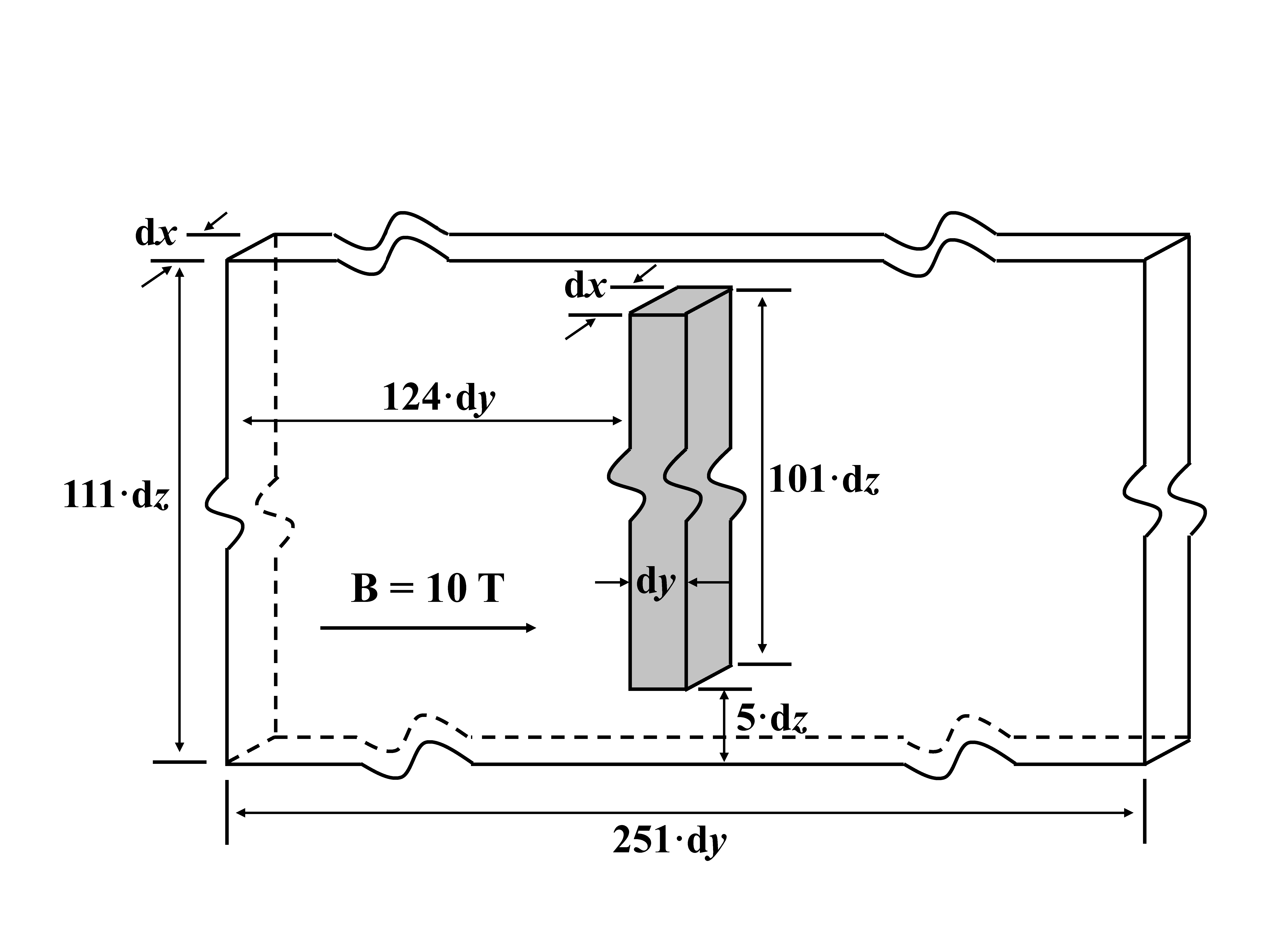}
\par\end{centering}
\caption{Initial particle distribution. The computational region was of size
$1\cdot\mathrm{d}x\times251\cdot\mathrm{d}y\times111\cdot\mathrm{d}z$
in JefiPIC and RGB-Maxwell, and of size $3\cdot\mathrm{d}x\times251\cdot\mathrm{d}y\times111\cdot\mathrm{d}z$
in EPOCH. The particles are placed in the shaded region uniformly
in the cuboid volume of $1\cdot\mathrm{d}x\times1\cdot\mathrm{d}y\times101\cdot\mathrm{d}z$
ranging from $0\cdot\mathrm{d}x$ to $1\cdot\mathrm{d}x$ ($1\cdot\mathrm{d}x$
to $2\cdot\mathrm{d}x$ in EPOCH) in the $x$-axis, $124\cdot\mathrm{d}y$
to $125\cdot\mathrm{d}y$ in the $y$-axis and $5\cdot\mathrm{d}z$
to $106\cdot\mathrm{d}z$ in the $z$-axis. The velocities of the
particles are zero.\label{fig:Initial-particle-distribution.-1}}
\end{figure}

As previously mentioned, the integral-based PIC method has the advantage of its ability to handle non-neutral plasma without Poisson pre-processing. Here, we compare the evolution of a batch of electrons with zero velocity in JefiPIC, EPOCH, and RGB-Maxwell. The total charge of electrons is $-5\times10^{-14}$ C. We select the computational region size as $n_{x}\cdot n_{y}\cdot n_{z}=1\times251\times111$ in JefiPIC and RGB-Maxwell, and $n_{x}\cdot n_{y}\cdot n_{z}=3\times251\times111$ in EPOCH. Initially, these electrons are placed uniformly within a cuboid region of size $1\cdot\mathrm{d}x\times1\cdot\mathrm{d}y\times101\cdot\mathrm{d}z$, ranging from $0\cdot\mathrm{d}x$ to $1\cdot\mathrm{d}x$ ($1\cdot\mathrm{d}x$ to $2\cdot\mathrm{d}x$ in EPOCH) in the $x$-axis, $124\cdot\mathrm{d}y$
to $125\cdot\mathrm{d}y$ in the $y$-axis and $5\cdot\mathrm{d}z$
to $106\cdot\mathrm{d}z$ in the $z$-axis, shown in Figure \ref{fig:Initial-particle-distribution.-1}. To better observe the evolution of particles, we apply a constant magnetic field of $10$ T along the y-direction to constrain the electrons' transverse motion, considering velocity only in the $y$-direction.

\begin{figure}
\begin{centering}
\includegraphics[scale=0.85]{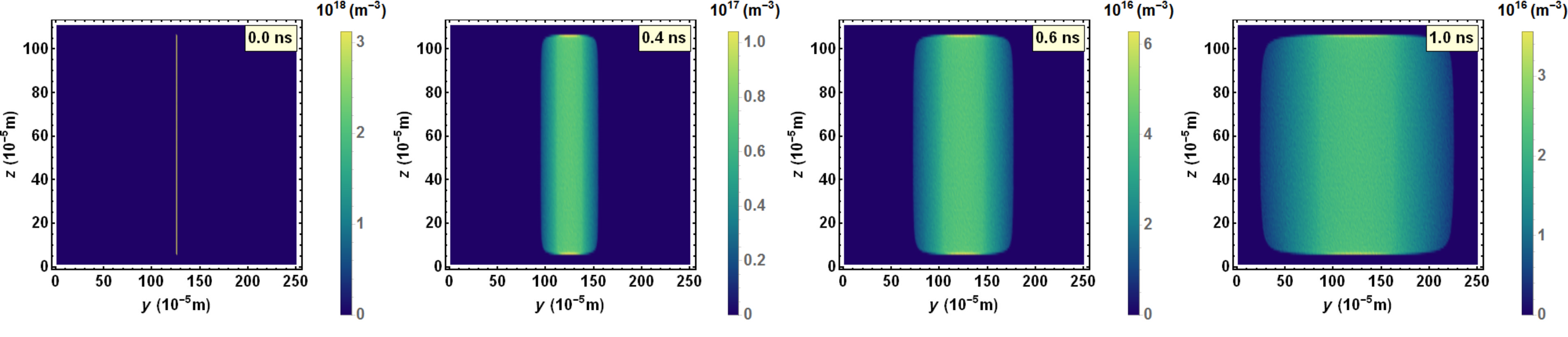}
\par\end{centering}
\begin{centering}
\includegraphics[scale=0.85]{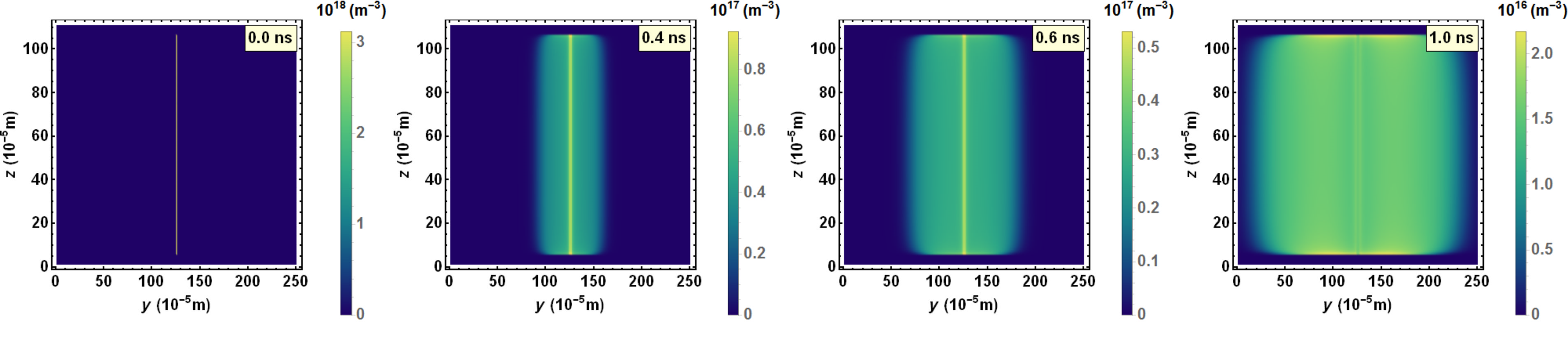}
\par\end{centering}
\begin{centering}
\includegraphics[scale=0.85]{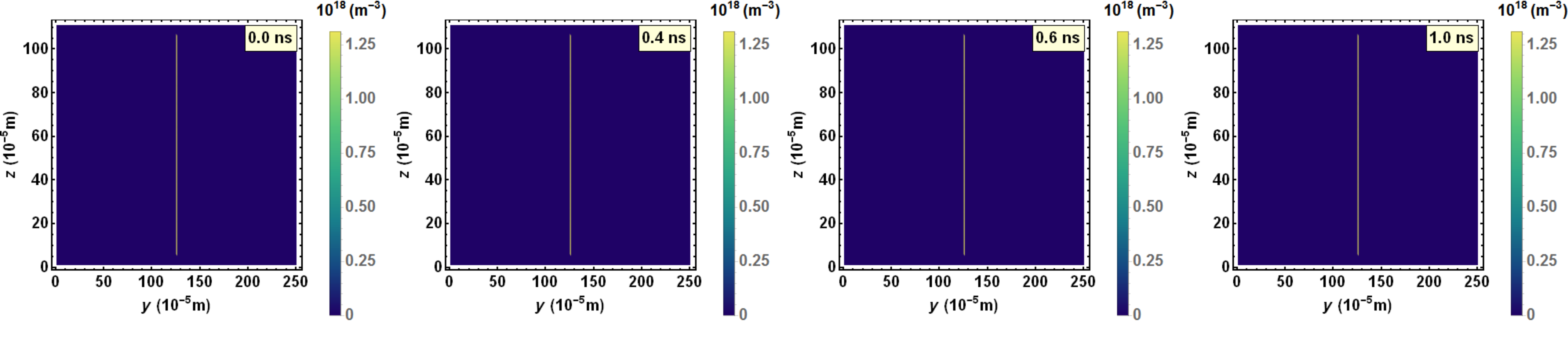}
\par\end{centering}
\caption{Particle distributions over time. The three rows represent the particle
distributions obtained from JefiPIC, RGB-Maxwell, and EPOCH, respectively.
Panel (a) shows the distribution at 0.0 ns, followed by (b) at 0.4
ns, (c) at 0.6 ns, and (d) at 1.0 ns.\label{fig:Particle-distributions-over-1}}
\end{figure}

Due to the mutual repulsion between particles, it is expected that
they should diffuse from the center to both sides along $y$-axis.
As depicted in Figure \ref{fig:Particle-distributions-over-1}, particles
simulated by JefiPIC and RGB-Maxwell consistently spread out as expected.
However, EPOCH fails to demonstrate a reasonable distribution when
pre-processing tasks, such as Poisson's equation, are not executed
correctly. This outcome can be explained by the fact that JefiPIC
and RGB-Maxwell utilize Jefimenko's equations, which can model the
EM field as long as charge and current densities are supplied. Therefore,
even in the absence of initial current, the electrostatic fields calculated
from the charge density can still drive the charges apart. In contrast,
EPOCH is a current-based PIC code. To achieve the same result, we
must perform the pre-processing to offer the equivalent electrostatic
fields, such as solving the Poisson's equation. However, this essential
operation will lead to additional computation time.

\begin{figure}
\begin{centering}
\includegraphics[scale=0.4]{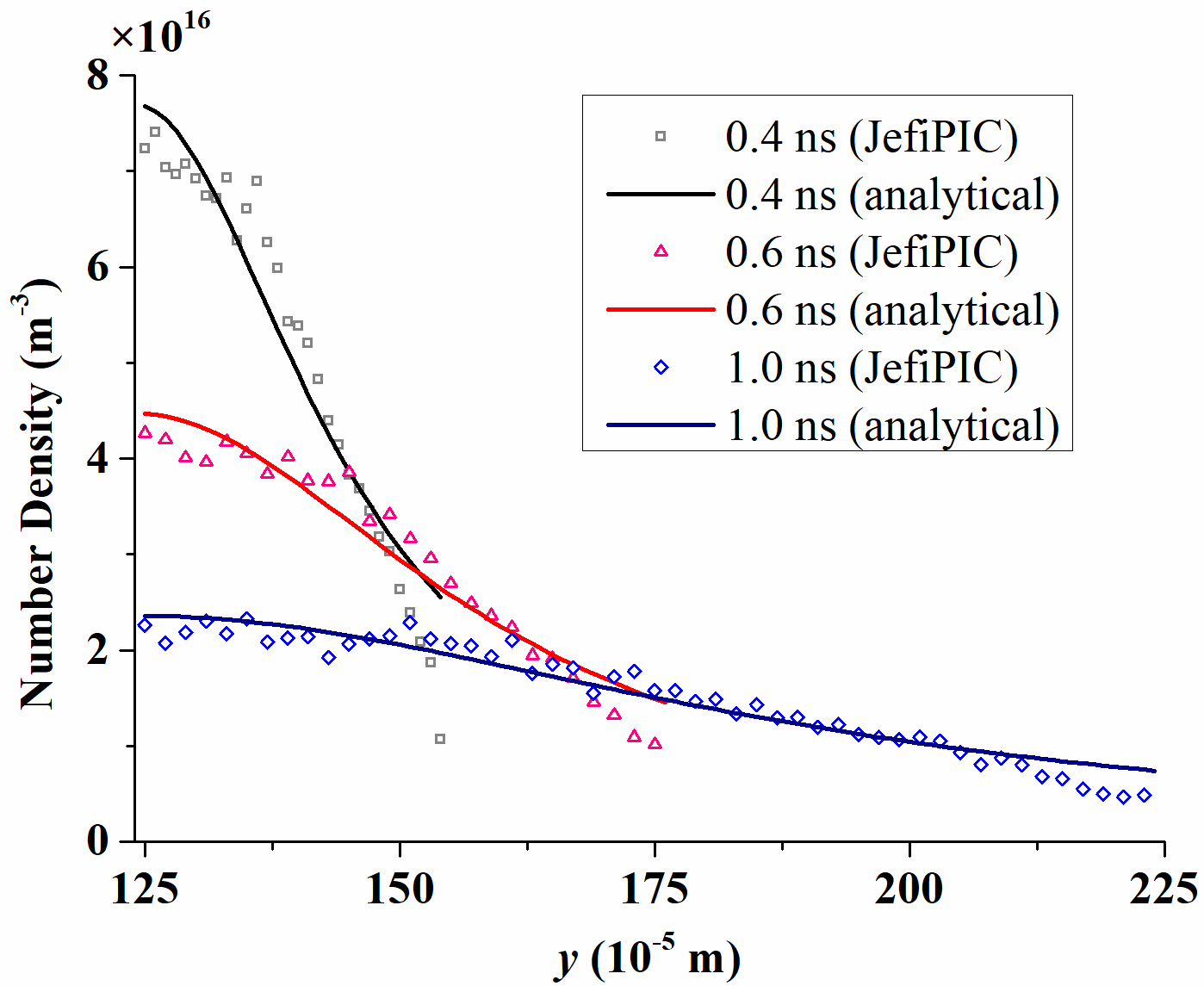}
\par\end{centering}
\caption{Simulated and analytical particle number density distribution. The
figure displays the simulated particle number density distribution
obtained through JefiPIC and the analytical solution from Reference
\cite{Lonngren-1976} at $t=0.4$ ns, 0.6 ns, and 1.0 ns. The data
are extracted from the grids ranging from {[}0 d$x$ to 1 d$x$, 125 d$y$
to 225 d$y$, 54 d$z$ to 55 d$z${]}. The gray hollow square, pink
hollow triangle and blue hollow rhombus represent the results by JefiPIC,
while the black, red, and dark blue lines represent the analytical
results.\label{fig:Simulated-and-analytical}}
\end{figure}

To further validate the accuracy of JefiPIC, we compare its particle
number density distribution with the analytical solution \cite{Lonngren-1976},
\begin{eqnarray}
\psi & \sim & \frac{1}{\tau\left[y^{2}/2\tau^{2}+1/\psi_{0}\right]},\label{eq:analytical_sol}
\end{eqnarray}
where $\psi$,$\,\tau$, and $y$ represent the normalized number
density, time and distance. As illustrated in Figure \ref{fig:Simulated-and-analytical},
the particle distributions produced by JefiPIC closely match the analytical
solution. 

Constrained by the requirement to adopt non-absorbing boundary conditions when solving Poisson equations, the available open-source PIC codes with built-in Poisson solvers may inadvertently generate undesired reflected fields, which could consequently alter the motion of the particles

\subsubsection{Two stream instability}
\begin{figure}
\begin{centering}
\includegraphics[scale=0.4]{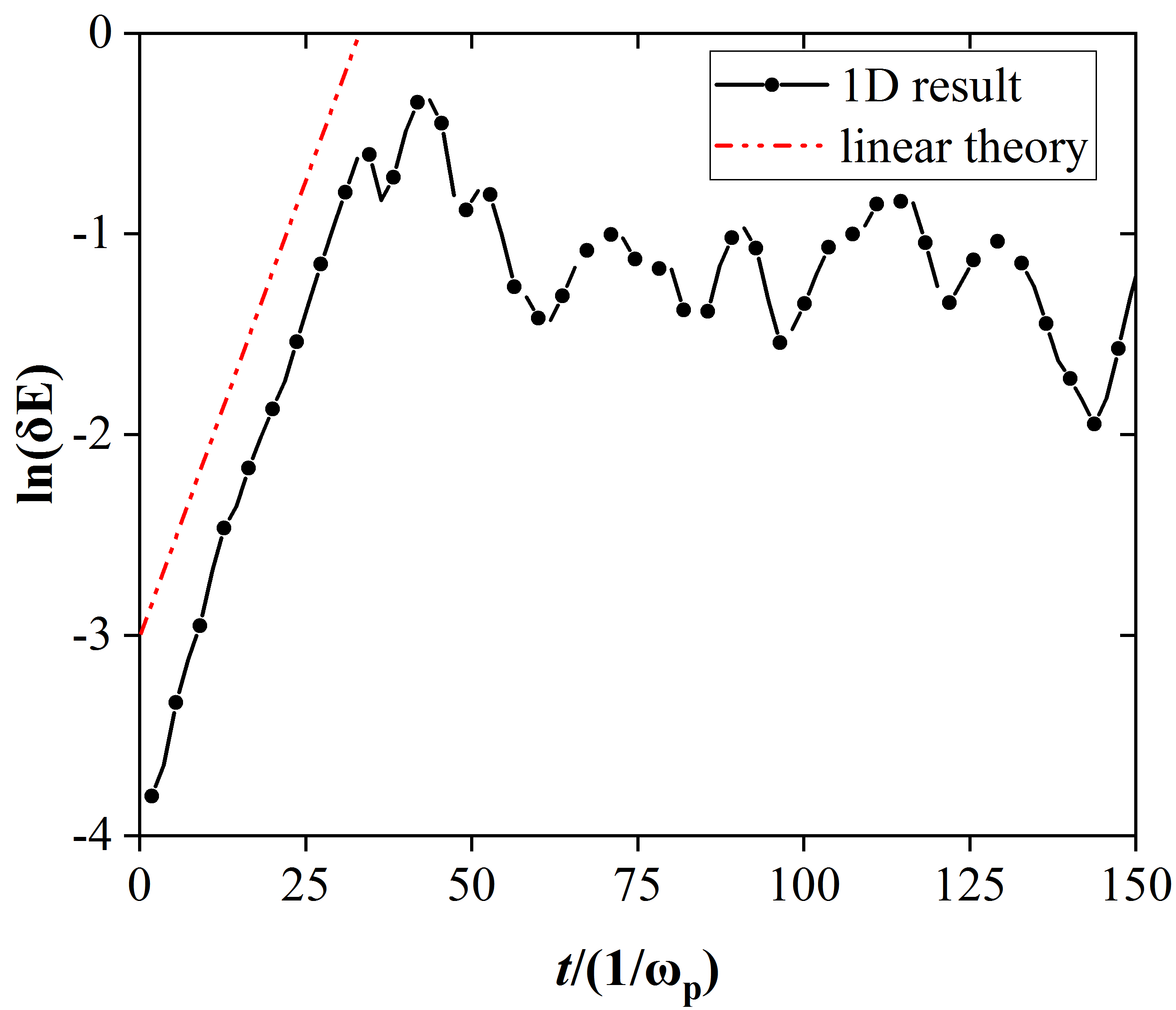}
\par\end{centering}
\caption{Growth of the mode with growth rate in the two stream instability. The black line represents the 1D numerical results, and the red line is the prediction of linear theory, which are consistent with each other. We set $L_{y} = 3\times10^{-4}$, $k=2\pi/L_{y}$ and $\gamma=0.0935$.\label{fig:two-stream}}
\end{figure}

\begin{figure}
\begin{centering}
\includegraphics[scale=0.3]{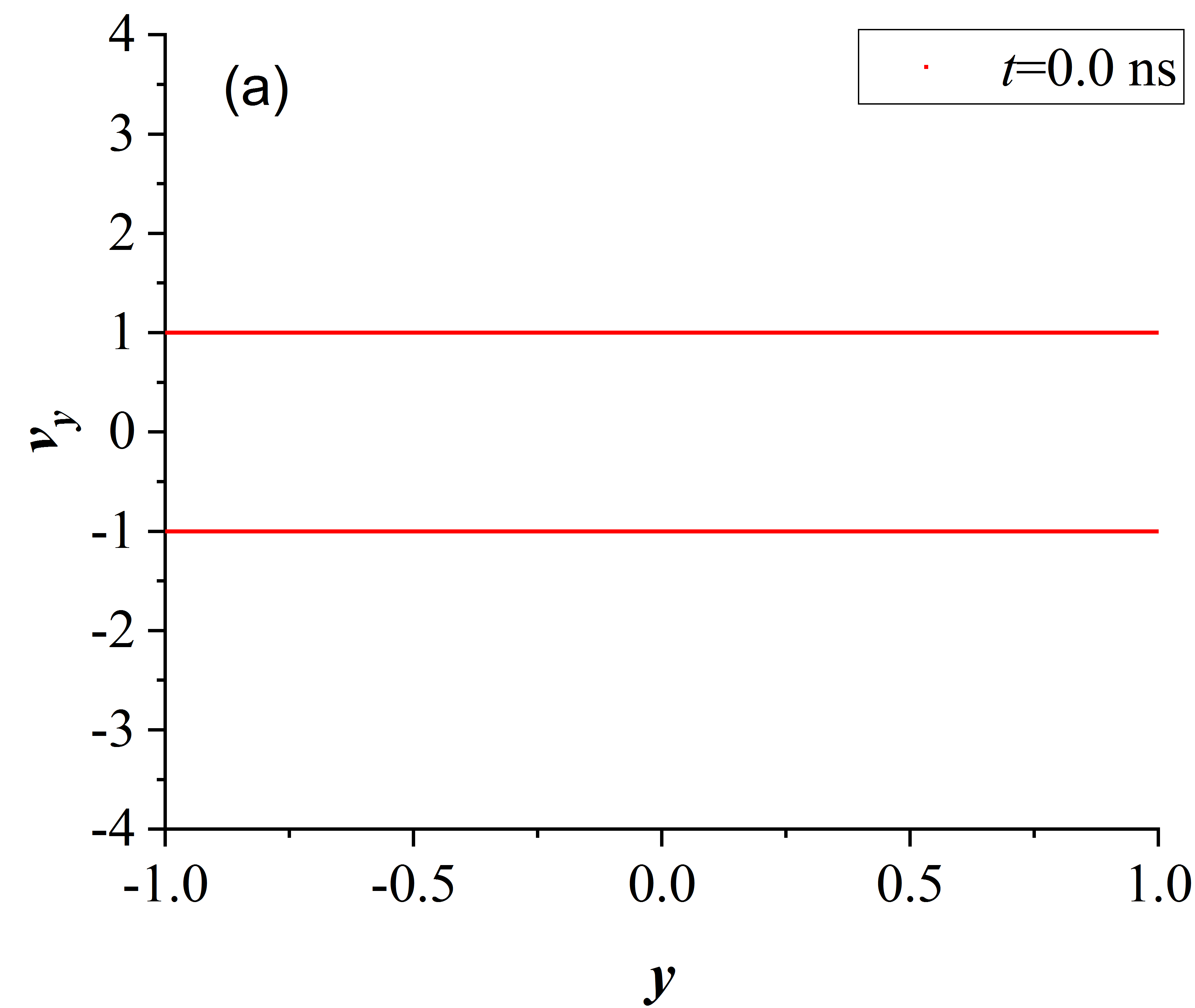}
\par\end{centering}
\begin{centering}
\includegraphics[scale=0.3]{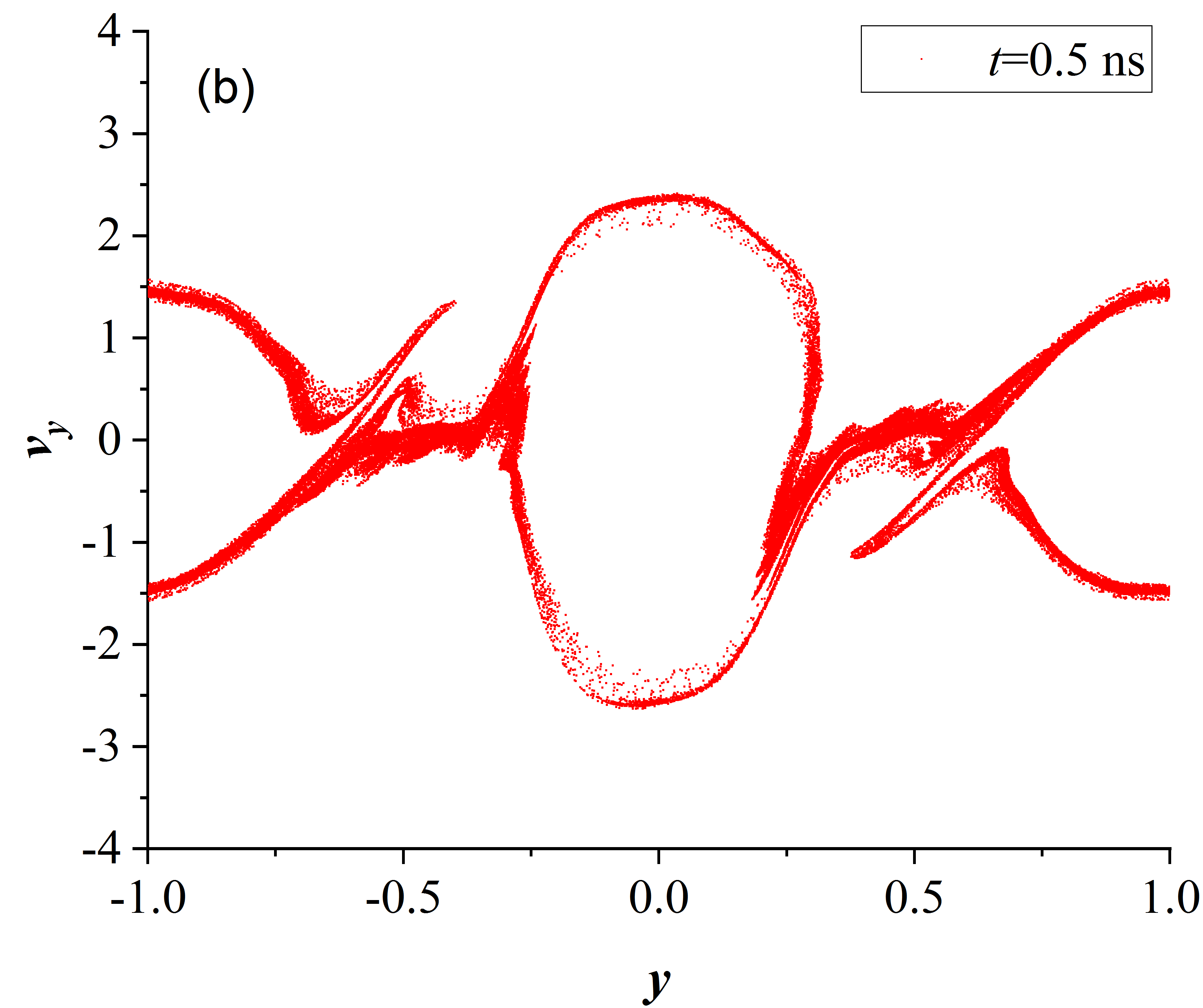}
\par\end{centering}
\begin{centering}
\includegraphics[scale=0.3]{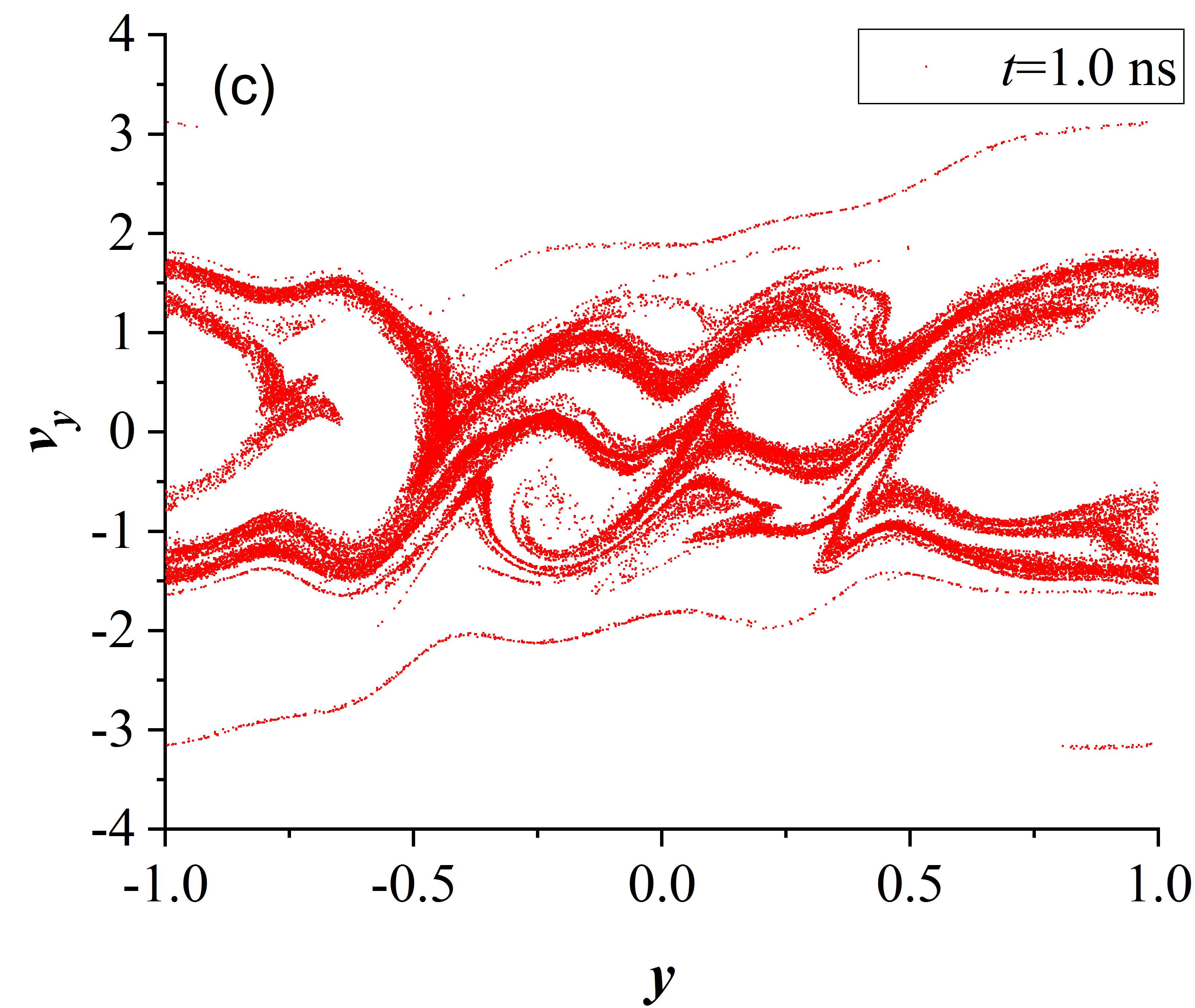}
\par\end{centering}
\caption{Particle phase space plots for the 1D two stream instability problem. (a) $t=0.0 ns$, (b) $t=0.5 ns$, (c) $t=1.0 ns$. The initial velocities and displacements are normalized.\label{fig:Particle-phase}}
\end{figure}

Applied with period boundary condition according to Ref.\cite{ewald-sum-2009}, we simulate the 1-D two stream instability with two counter-streaming electron beams with initial energy of $E_{0}$ = 1eV along $y$ axis. The two electron beams consisting of total charge $-5\times10^{-14}$ C are emitted into a static uniform neutralizing background charge. We take $L_{y} = 3\times10^{-4}$ m and $k=2\pi/L_{y}$, the computational time $t = 2 ns$ and the time step $\mathrm{d}t = 4\times10^{-4}$ ns. According to the dispersion relation for the two stream instability from Ref.\cite{two-streamn-2014}, we have
 
\begin{equation}
D\left(\omega,k\right)=1-\omega_{p}^{2}\left[\frac{1}{\left(\omega-ku\right)^{2}}+\frac{1}{\left(\omega+ku\right)^{2}}\right]
\end{equation}
which gives the greatest growth rate of $\gamma=0.0935$, where $\omega_{p}$ is the plasma frequency, and $u$ is the relativistic velocity of particle. The growth of this mode of electric field (black line) is shown in Figure \ref{fig:two-stream}, and agrees well with the rate from linear theory (red line). We also plot the phase space distribution in Figure \ref{fig:Particle-phase}, and the two separate distribution lines “roll up” obviously after the interaction of beams.

\subsection{Obeying Charge Conservation Law}
Charge conservation is indeed critical in PIC simulation. The linear weightings for current may not satisfy the continuity equation in the finite difference method. This is because the finite difference-based PIC method implicitly computes electric field solely through electric current. Although theoretically, as long as current continuity is established, the conservation property is preserved, the linear current interpolation method based on the finite difference method can potentially violate the conservation law, thus causing the electric field divergence equation to fail.
However, by utilizing Jefimenko’s equation, a Green's function-like formulation, the electric field is explicitly computed by the actual sources - i.e., all the charge density, time-varying charge density, and current density, and integrated across the entire computational region. This integration method can circumvent the charge conservation problem.

\subsubsection{Verification of charge conservation law}
\begin{figure}
\begin{centering}
\includegraphics[scale=1]{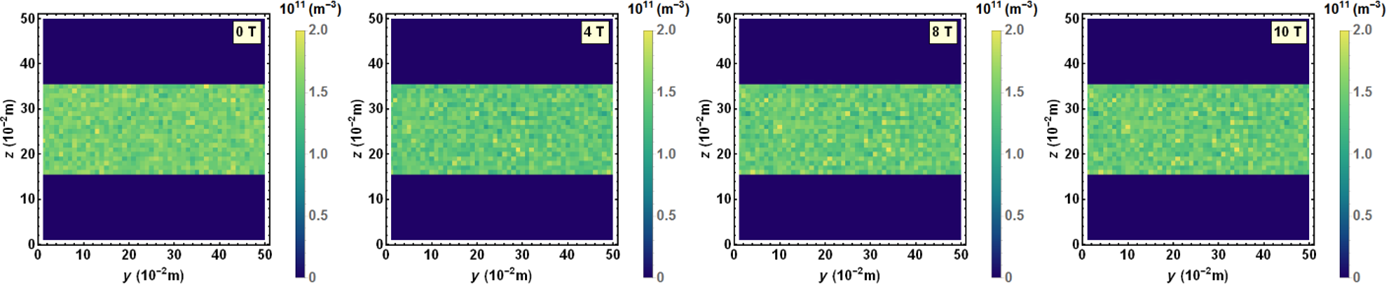}
\par\end{centering}
\caption{Beam distribution at different time shots. T represents the electron transit period for electron initial energy of 30 keV and the transit distance long $y$ axis of 1 m.\label{fig:Beam}}
\end{figure}

We employ JefiPIC to simulate the beam expansion problem proposed in Ref. \cite{charge-conservation-1997} to verify the charge conservation law. Electrons with initial energy of 30 keV and a current of 5 A are injected from the left boundary into the space and are removed from the simulation as long as they arrive the right boundary. The simulation space adopts the absorbing field and particle boundary condition. Figure \ref{fig:Beam} shows the particle distributions simulated by our PIC code which only uses a simple bilinear current weighting without using a divergence correction. The result shows that the beam spreads transversely only slightly as it traverses the system due to its space charge force during the simulation up to 10 transit period.

\subsubsection{Compared with UNIPIC}
\begin{figure}
\begin{centering}
\includegraphics[scale=0.4]{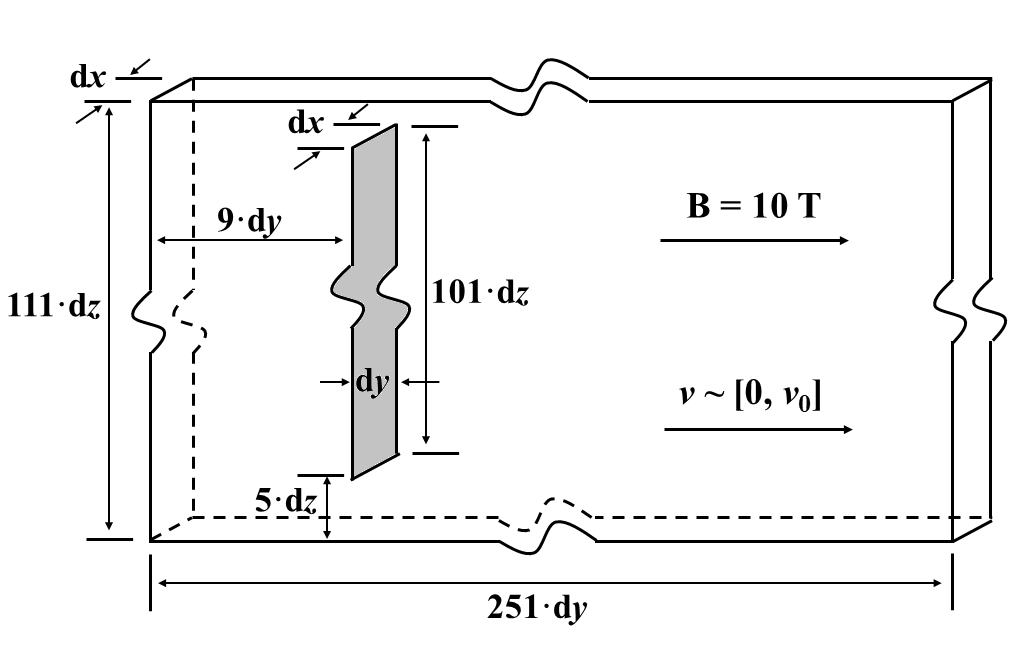}
\par\end{centering}
\caption{Initial electron distribution in the simulation. The total computational
region was of size 1 d$x$\texttimes 251 d$y$\texttimes 111 d$z$
in JefiPIC and 3 d$x$\texttimes 251 d$y$\texttimes 111 d$z$ in
UNIPIC. The particles were placed uniformly on the shaded rectangle
face of $y=9\cdot\mathrm{d}y$ and 1 d$x$\texttimes 101 d$z$, with
0 d$x$ to 1 d$x$ (1 d$x$ to 2 d$x$ in UNIPIC) in the $x$-axis
and 5 d$z$ to 106 d$z$ in the $z$-axis. Particle velocities obeyed
a random distribution from 0 to $v_{0}$.\label{fig:Initial-electron-distribution}}
\end{figure}

We choose another FDTD-based PIC code, UNIPIC, to further demonstrate the advantage of the integral-based PIC codes when dealing with the charge-conservation problem. Since UNIPIC obeys the charge-conservation law that when electrons are emitted from the shaded rectangle face, the pseudo positive charge with the opposite charge will retain at the emission surface. The total charge of electrons is $-5\times10^{-14}$ C. The electrons' initial velocities in the $y$-direction are randomly from 0 to $v_{0}$. We select the computational region size as $n_{x}\cdot n_{y}\cdot n_{z}=1\times251\times111$ in JefiPIC, and $n_{x}\cdot n_{y}\cdot n_{z}=3\times251\times111$ in UNIPIC. We apply a constant magnetic field of $10$ T in the y-direction again to constrain the electrons' transverse motion, shown in Figure \ref{fig:Initial-electron-distribution}. Thus, only velocities in the $y$-direction are considered.

\emph{Note that: UNIPIC employs the simple linear interpolation method rather than the strict charge-conservation method, and thus it needs to use the Langdon-Marder correction algorithm at a certain frequency.}

\begin{figure}
\begin{centering}
\includegraphics[scale=0.7]{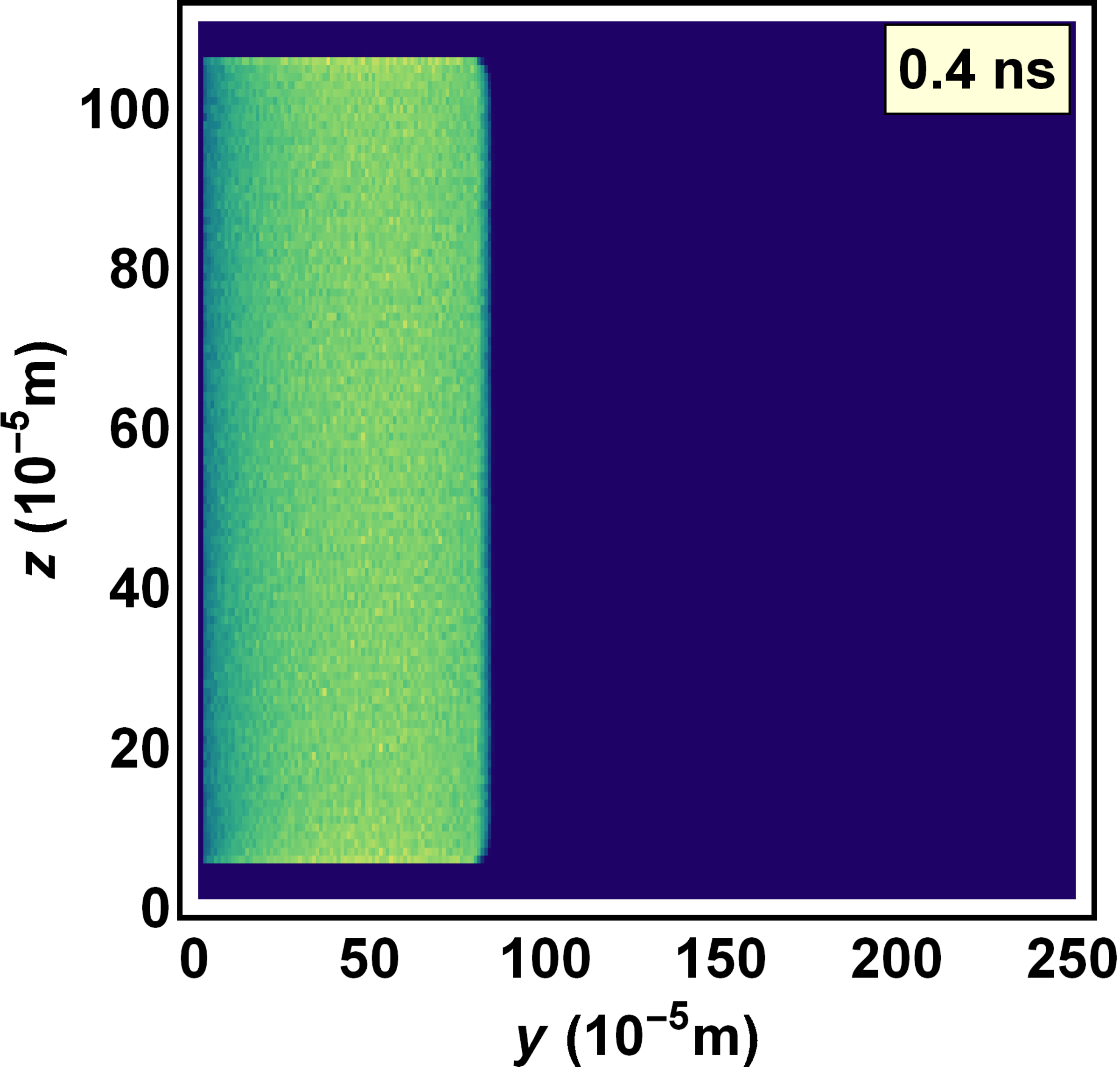}\includegraphics[scale=0.7]{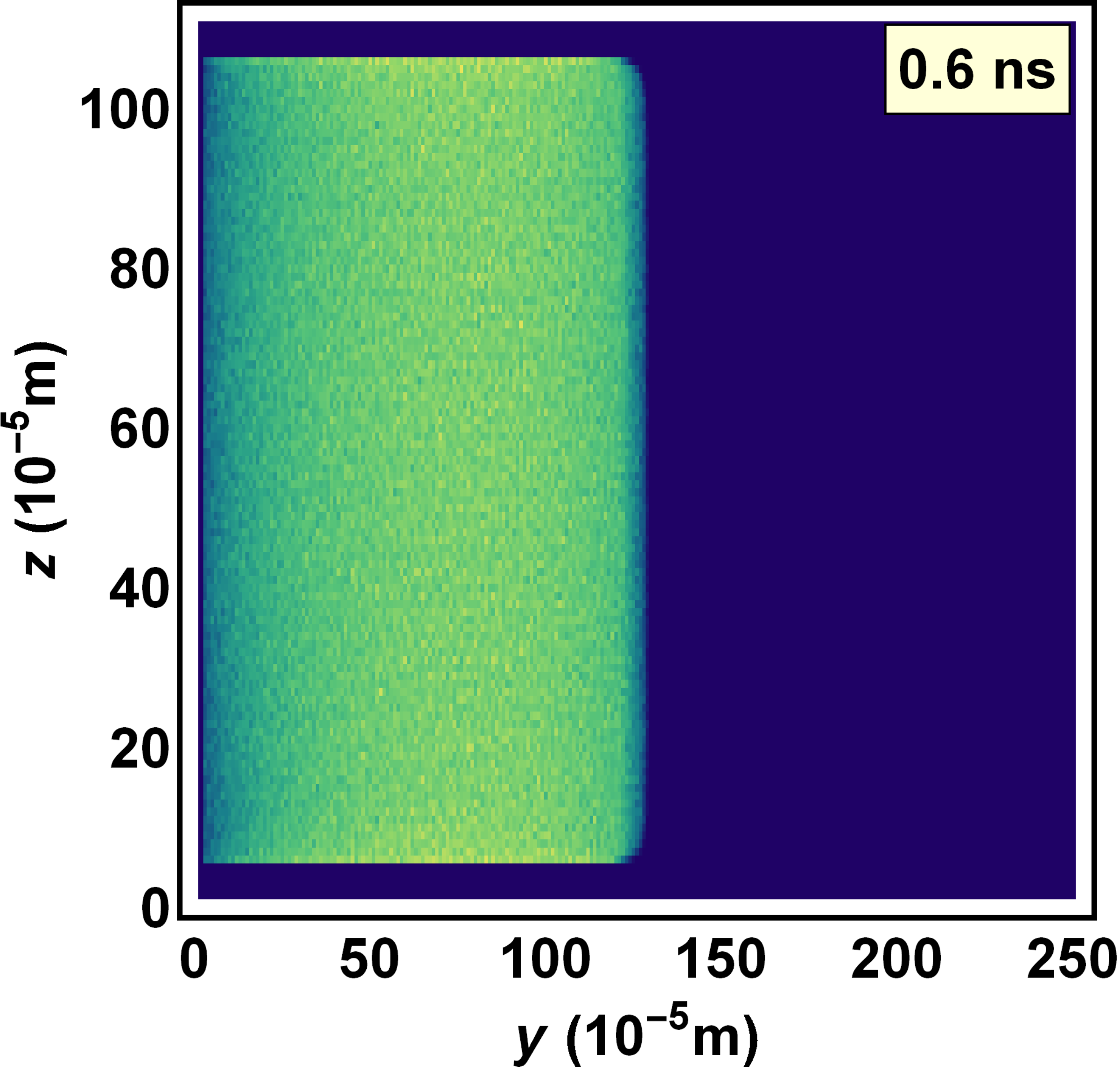}
\par\end{centering}
\begin{centering}
\includegraphics[scale=0.22]{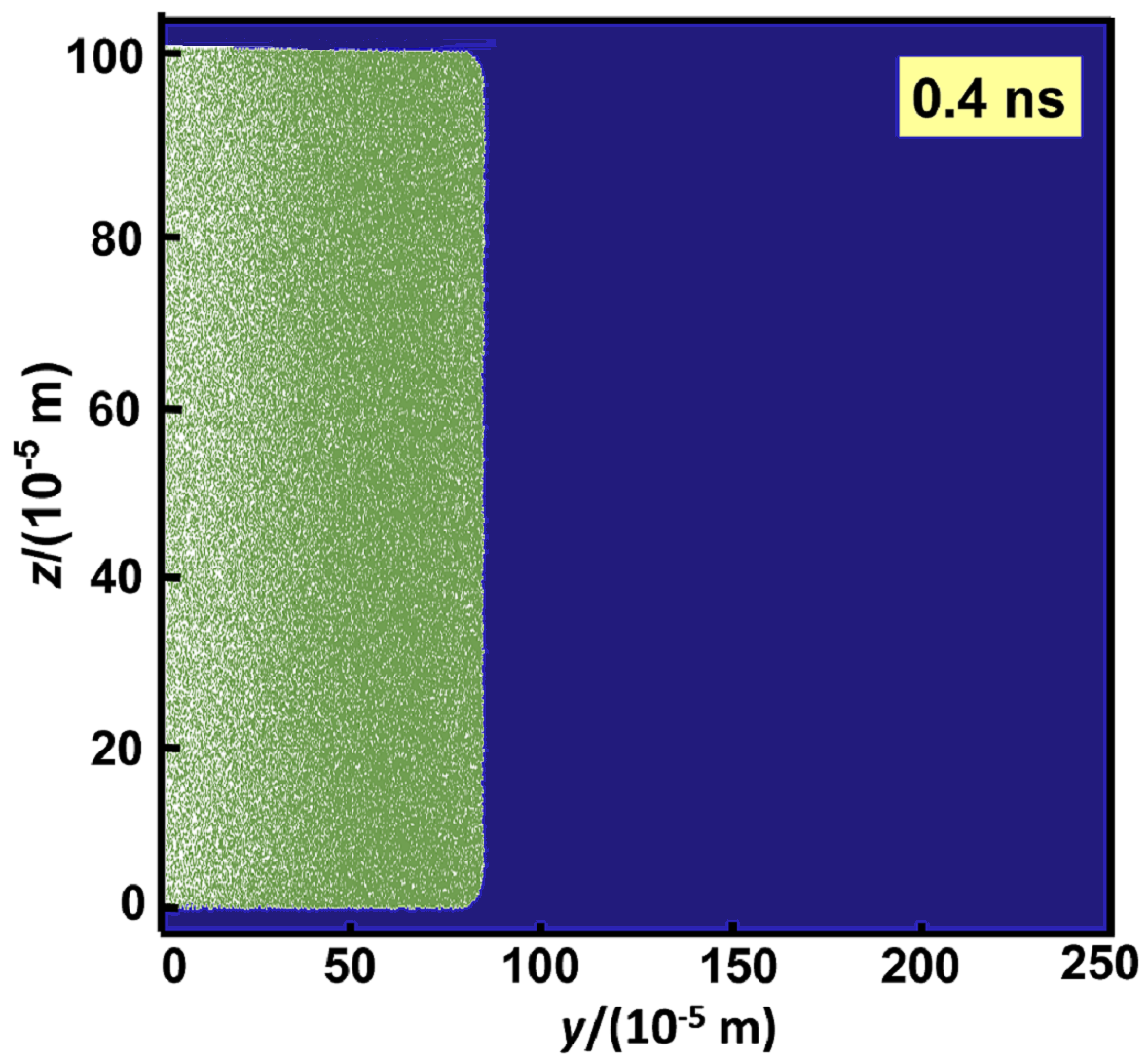}\includegraphics[scale=0.26]{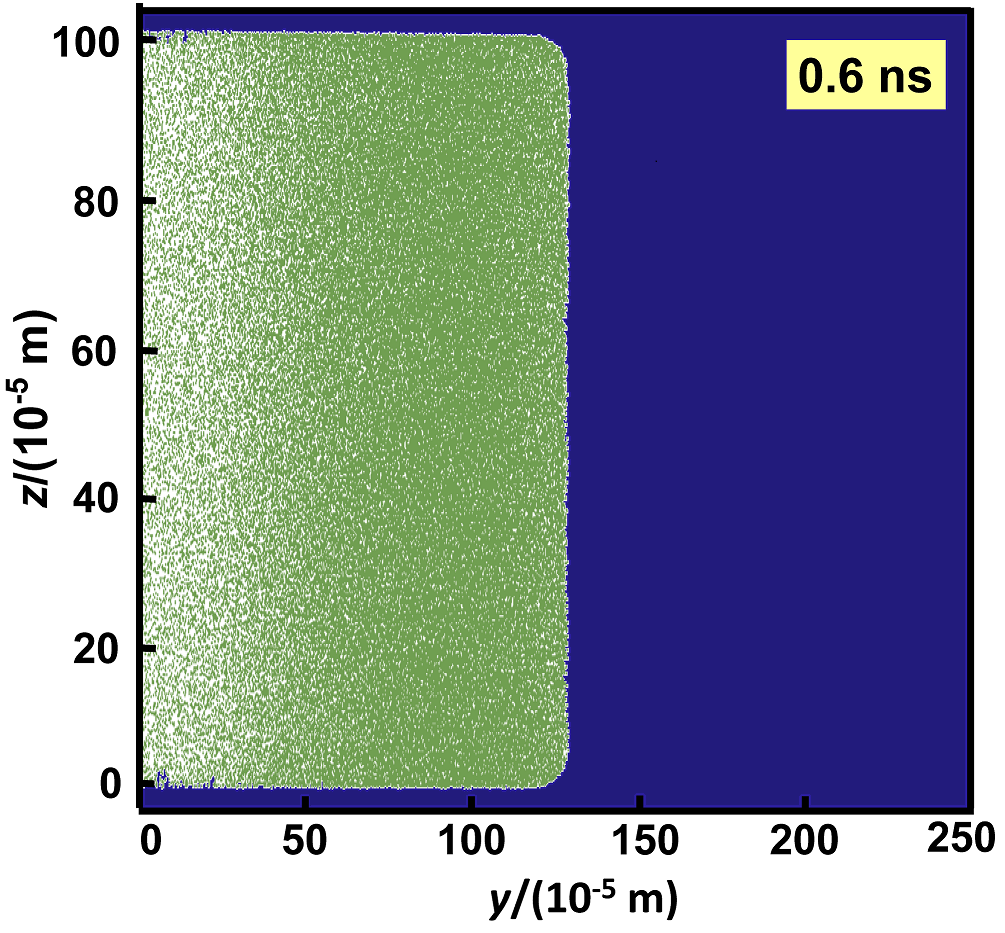}
\par\end{centering}
\caption{Comparison of particle distributions at t = 0.4 ns and t = 0.6 ns
in (a) JefiPIC and (b) UNIPIC simulations. Initial velocities were
applied only in the $y$-direction, randomly ranging from 0 to $v_{0}$.
This illustrates the differences in Coulomb forces between the two
simulations, highlighting the more accuracy of JefiPIC in simulating
particle movements.\label{fig:Comparison-of-particle}}
\end{figure}

Figure \ref{fig:Comparison-of-particle} displays the particle distribution obtained from JefiPIC and UNIPIC at two different times. A comparison of the two distributions reveals their overall similarity, validating our code's results to a certain extent. Table \ref{tab:Comparison-of-the} lists the maximum displacement and velocity of particles under corresponding time steps in Figure \ref{fig:Comparison-of-particle}. It is shown that the particles states simulated by JefiPIC and UNIPIC are in close agreement, with deviation around $1\sim3\%$.
\begin{table}
\caption{Comparison of the maximum particle displacement and velocity between
JefiPIC and UNIPIC\label{tab:Comparison-of-the}}

\centering{}%
\begin{tabular}{c|c|c|c|c}
\hline 
\multirow{2}{*}{Time/ns} & \multicolumn{2}{c|}{Maximum Displacement/m} & \multicolumn{2}{c}{Maximum Velocity/(m s-1)}\tabularnewline
\cline{2-5} \cline{3-5} \cline{4-5} \cline{5-5} 
 & JefiPIC & UNIPIC & JefiPIC & UNIPIC\tabularnewline
\hline 
0.4 & $8.43\times10^{-4}$ & $8.49\times10^{-4}$ & $2.20\times10^{6}$ & $2.26\times10^{6}$\tabularnewline
\hline 
0.6 & $1.29\times10^{-3}$ & $1.30\times10^{-3}$ & $2.25\times10^{6}$ & $2.30\times10^{6}$\tabularnewline
\hline 
\end{tabular}
\end{table}

\begin{figure}
\begin{centering}
\includegraphics[scale=0.4]{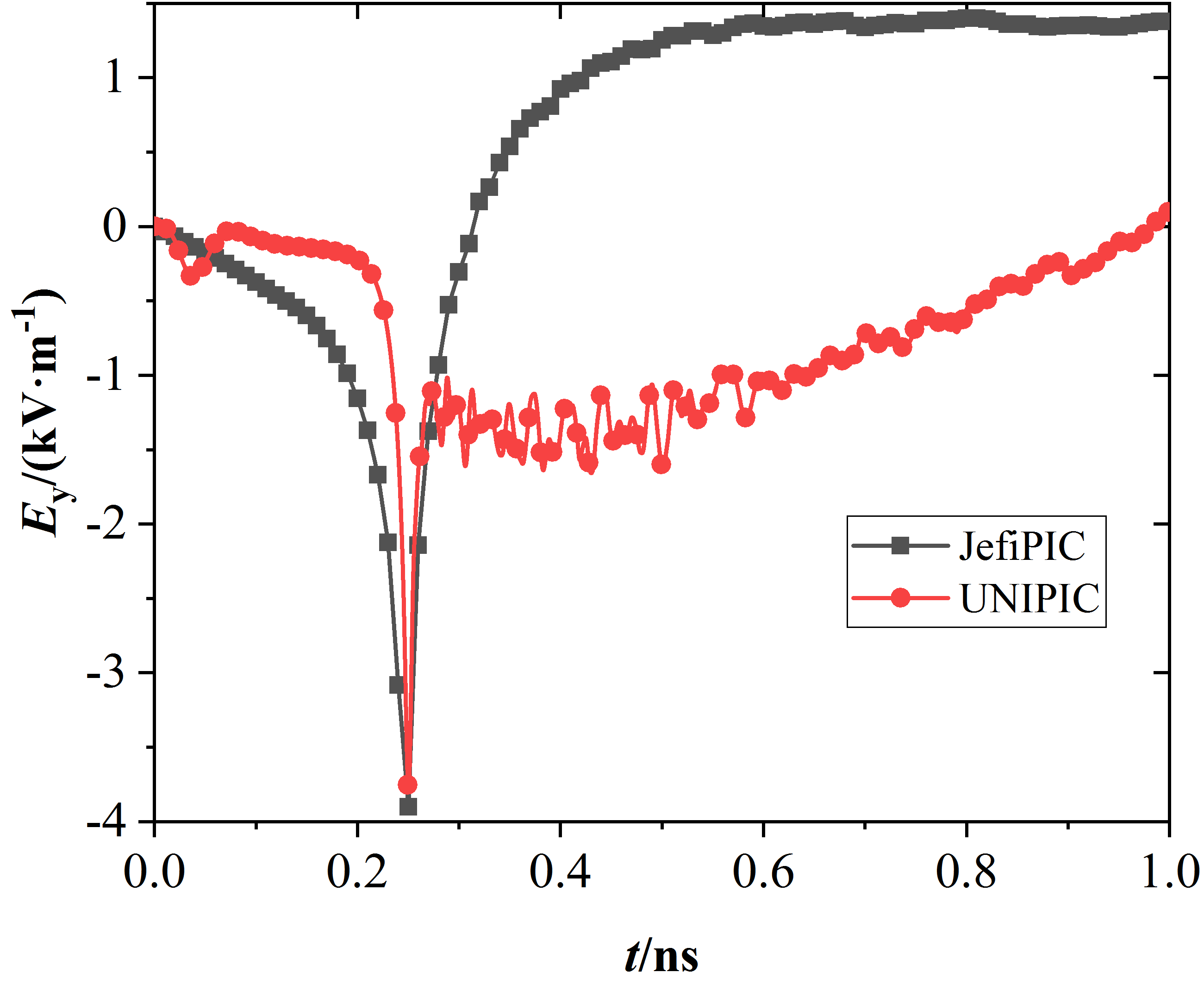}
\par\end{centering}
\caption{Comparison of $y$-direction electric field. The $y$-direction electric
fields at point $(i=0,j=50,k=55)$ derived from JefiPIC and at point
$(i=1,j=50,k=55)$ from UNIPIC are compared. The black square and
red circle represent the results from JefiPIC and UNIPIC.\label{fig:Comparison-of-y-direction-1}}
\end{figure}

Figure \ref{fig:Comparison-of-y-direction-1} exhibits the y-direction electric fields at point $(i=0,j=50,k=55)$ in JefiPIC and at point $(i=1,j=50,k=55)$ in UNIPIC for detail. The electric fields of the two simulations match with each other in the respect of peak value and peak time, but differ in two aspects: a) the pulse width of JefiPIC's field is larger, and b) JefiPIC's trailing edge field changes to positive due to the charge separation. Besides, JefiPIC shows a significantly smoother electric field profile in comparison to UNIPIC. 

Thus, in the aspect of charge conservation, the integral method provides a natural advantage in terms of accuracy and simplicity over the difference method in PIC, despite the fact that the difference-PIC has already incorporated the necessary Langdon-Marder amendment algorithms.

We also compared the computational performance of the two codes. UNIPIC
was run on a workstation with Intel Xeon Processor E5-2650 v4 CPU across
200 threads, and it took over 6 hours to complete the calculations.
In contrast, using a larger time step, JefiPIC completed the computations
within only half an hour on a single GPU A100 card, achieving over
90\% time-saving benefits.

\subsection{Choice of Macro-Particle, Grid number, Time step and $N_{\mathrm{d}t}$}

\begin{figure}
\begin{centering}
\includegraphics[scale=0.4]{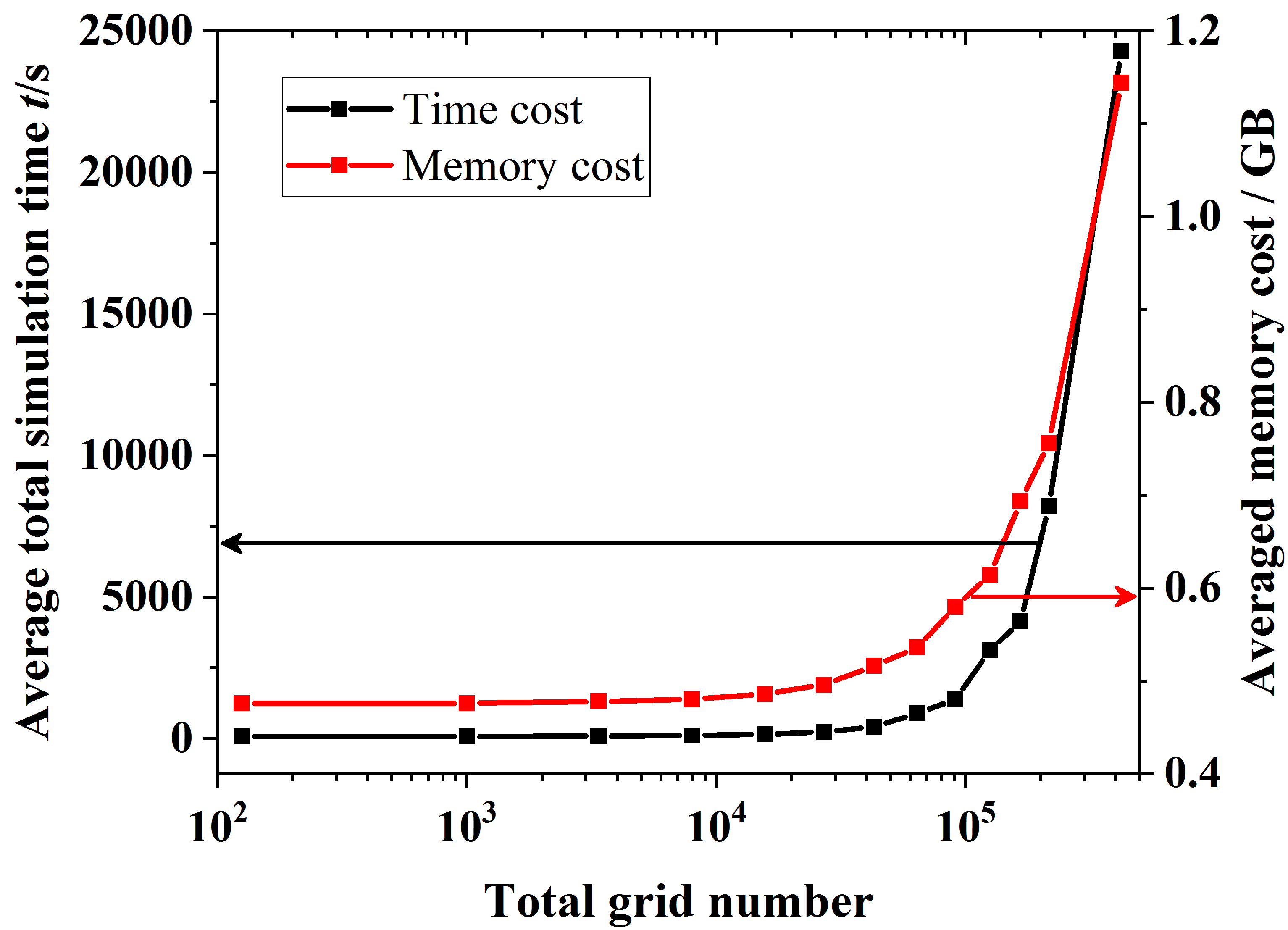}
\par\end{centering}
\begin{centering}
\includegraphics[scale=0.4]{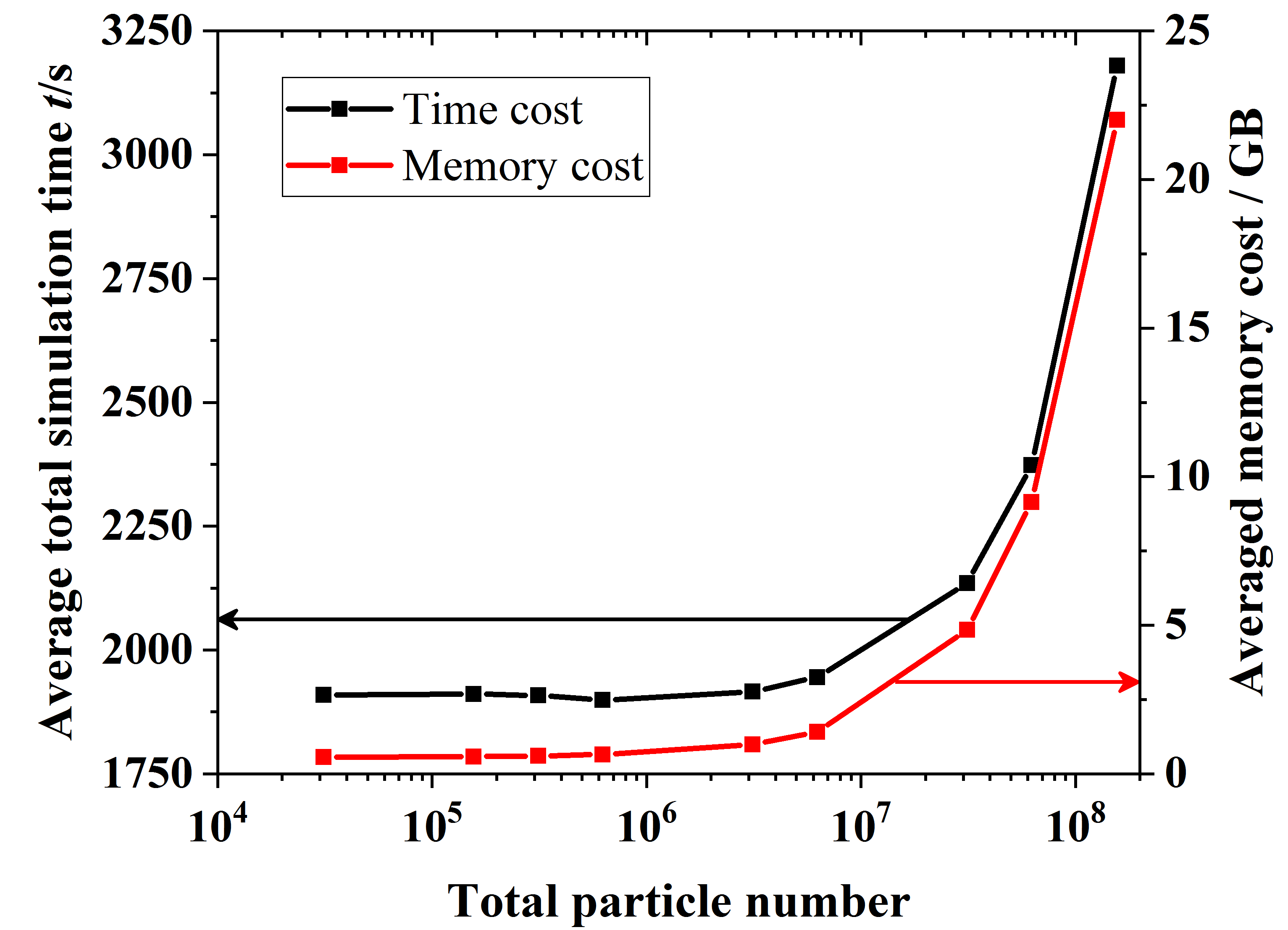}
\par\end{centering}
\caption{Time cost of JefiPIC with varying particle numbers. The maximum CUDA
threads limits the maximum number of particles to $10^{9}$.\label{fig:Time-cost-of}}
\end{figure}

We conduct test on JefiPIC's time consumption with varying macro-particle
numbers and grid numbers to provide readers with additional reference. It can be concluded
from Figure \ref{fig:Time-cost-of} (a) that when the particle number
is less than one percent of the CUDA kernel limitation ($\sim$just
under $10^{7}$), the computational time keeps relatively short due
to the high degree of parallelism on GPU and a small number of data
transfer between CPU and GPU. However, once the particle number exceeds
$10^{7}$, the time cost increase significantly. Note that the particle
number in JefiPIC (i.e., the occupied kernel number) must not exceed
the CUDA limitation ($~10^{9}$), or the computation will fail. Fortunately,
$10^{7}\sim10^{8}$ macro-particles are typically sufficient for most
simulations within the controlled noises.

Besides, the FLOPS of the A100 card is approximately 312 TFLOPS, and the efficiency of our code on A100 exceeds 99\%, implying that the FLOPS of JefiPIC is also around 312 TFLOPS on A100. We count the computational time the three main kernels cost, shown in Table  2. The results indicate that the interpolation process occupies the majority of computational time (around over 95\%), and that total I/O operations require approximately 60 seconds, which is relatively negligible when compared to the extensive computational time executing the CUDA kernel functions. Increasing the number of grids will extend interpolation time, while augmenting the number of particles will increase particle motion time.

As for the grid number exhibited in Figure \ref{fig:Time-cost-of} (b), the similar tendency can be observed. In the grid range we compute, though the GPU memory cost increase only no more than three times with the increase of grid number, the time cost increase over 4000 times when the grid number exceeds around $10^{5}$, while on the contrary the increase rate of memory cost is larger for the increase of macro-particle. We find that JefiPIC is able to solve a project with maximum of $10^{7}$ grid on one GPU card limited by the GPU memory (40GB). However, the above condition is not achievable since the time cost is far too large.

\begin{table}
\caption{Comparison of the maximum particle displacement and velocity between
JefiPIC and UNIPIC\label{tab:Comparison-of-the}}

\centering{}%
\begin{tabular}{c|c|c|c|c}
\hline 
\multirow{2}{*}{Grid Size} & \multirow{2}{*}{Particle Number} & \multicolumn{3}{c}{Time cost/s (Proportion)}\tabularnewline
\cline{3-5} \cline{4-5} \cline{5-5} 
 &  & EM solver & Interpolation & Particle motion\tabularnewline
\hline 
\multirow{2}{*}{1\texttimes 251\texttimes 111} & $4.16\times10^{4}$ & 3.77 (1.05\%) & 476.74 (98.32\%) & 1.17 (0.63\%)\tabularnewline
\cline{2-5} \cline{3-5} \cline{4-5} \cline{5-5} 
 & $4.16\times10^{6}$ & 4.96 (0.78\%) & 465.75 (98.97\%) & 3.01 (0.25\%)\tabularnewline
\hline 
\multirow{2}{*}{1\texttimes 101\texttimes 101} & $4.16\times10^{4}$ & 3.74 (2.54\%) & 142.39 (96.68\%) & 1.16 (0.78\%)\tabularnewline
\cline{2-5} \cline{3-5} \cline{4-5} \cline{5-5} 
 & $4.16\times10^{6}$ & 5.64 (3.81\%) & 139.31 (94.25\%) & 2.86 (1.94\%)\tabularnewline
\hline 
\end{tabular}
\end{table}

\begin{figure}
\begin{centering}
\includegraphics[scale=0.4]{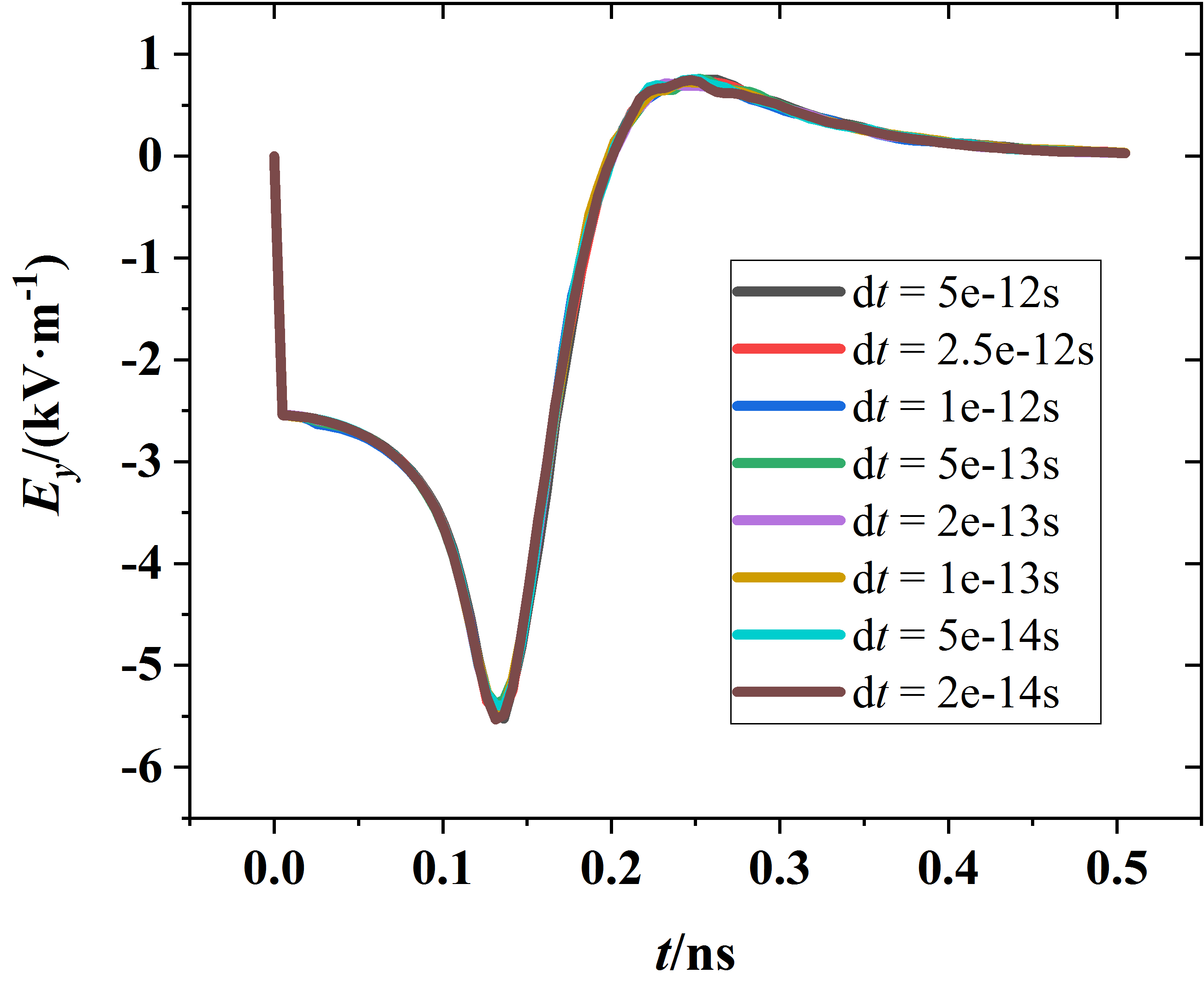}
\par\end{centering}
\caption{Electric fields with different time steps from CFL-limited time step ($2\times10^{-14}$ s) to its 250 times ($5\times10^{-12}$ s).\label{fig:dt}}
\end{figure}

Since we mentioned before that JefiPIC is not sensitive to the choice of time step when computing the EM field, we compare the electric field in the model of the circle expansion model in section 3.1. In this model, though the time step varies from the CFL-limited time step ($2\times10^{-14}$ s) to its 250 times ($5\times10^{-12}$ s), the electric field changes little, as  shown in Figure \ref{fig:dt}. We can conclude that the JefiPIC indeed has the advantages of the flexibility of the choice of time step.

\emph{Note that: The time step should also be limited by resolving the plasma period obtained according to the plasma density.}

Last, we report the influence of the parameter $N_{\text{d}t}$. In
our methodology, we indeed use a truncation level of 10000 to account
for the time history. However, the actual $N_{\text{d}t}$ required
to store the sources is determined by the minimum value between 10000
and the ratio of the diagonal length of the computational region $L_{\text{max}}$
to the time step dt used in Jefimenko's equation. Consequently, $L_{\text{max}}/(c\text{d}t)$
is frequently much smaller than 10000 and varies with the size of
the computational region. With this approach, in most scenarios, all
sources can be stored in the GPU to compute the fields, thus ensuring
the conservation properties of our method. The cases where the ratio
of $L_{\text{max}}/(c\text{d}t)$ exceeds 10000 generally correspond
to unrealistic scenarios with very large computational regions or
very small time steps. For the former, the influence of distant fields
on the observation point would be minimal. For the latter, it negates
the advantage of JefiPIC, which is the ability to use larger time
steps. In light of this, we have conducted an additional example where
the electric current density and electric charge density are governed
by analytical equations:

\begin{eqnarray}
J_{x}=J_{y}=J_{z}=\mathrm{sin}(x+y+z)\cdot\mathrm{sin}(t)\label{eq:J}
\end{eqnarray}

\begin{eqnarray}
\rho = 3(\text{cos}(t)-1)\cdot\text{cos}(x+y+z)\label{eq:rho}
\end{eqnarray}

\begin{figure}
\begin{centering}
\includegraphics[scale=0.4]{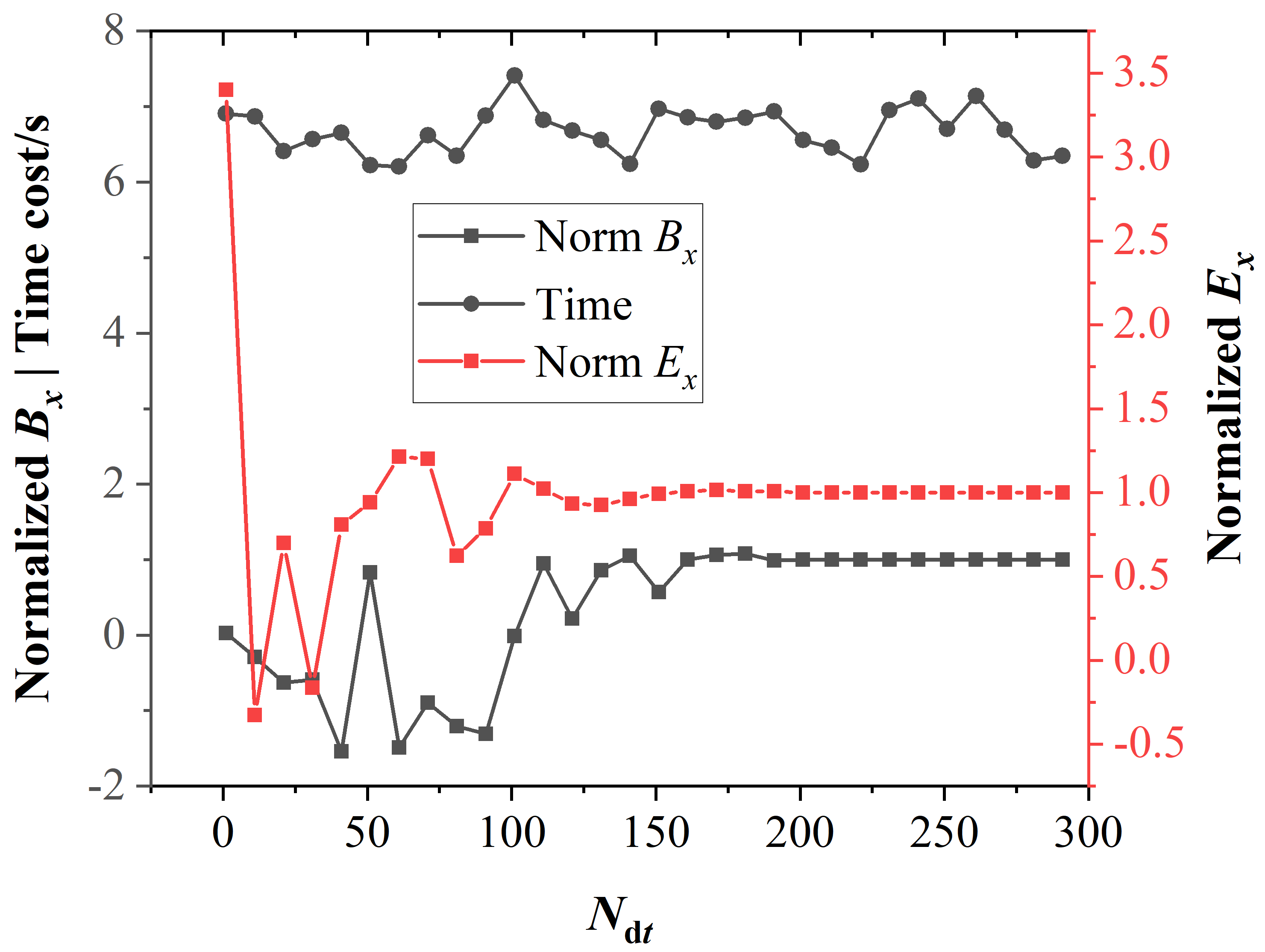}
\par\end{centering}
\caption{Normalized summation of the Magnetic Field $B_{x}$ and Electric Field $E_{x}$ over the entire space and variation of execution time with the change of $N_{dt}$.\label{fig:E-Ndt}}
\end{figure}

The above equations naturally satisfy the continuum equation. For
convenience, we set the fundamental constants $c$ (the speed of light),
$\epsilon_{0}$ (vacuum permeability) and $\mu_{0}$ (vacuum permittivity)
to be unit 1, i.e., the Rationalized Heaviside-Lorentz Units. We choose
a cubic of grid sizes $[20,20,20]$, with [d$x$,d$y$,d$z$]=[0.3,0.3,0.3] and d$t$=0.05. In the current set-up, the maximum
distance within the cubic is its diagonal, which is about $\sqrt{\text{3\texttimes\ensuremath{\left(20\times0.3\right)^{2}}}}$. Hence
the maximum time steps required for the electromagnetic wave to transmit
from one corner to another is $\sqrt{\text{3\texttimes\ensuremath{\left(20\times0.3\right)^{2}}}}/dt\approx207$, which means the maximum length of
the time steps tracked in the history should be larger than 207. To
see this, we loop through a series of track length, i.e., $N_{\text{d}t}=[1,11,21,...,291]$.
In Figure \ref{fig:E-Ndt}, we depict the scanned results. As can be seen, the calculated
electromagnetic fields display significant fluctuations when $N_{\text{d}t}$
is approximately less than 200. In contrast, when $N_{\text{d}t}$
is around 200 or more, the results appear to be consistent. This suggests
that the value of $N_{\text{d}t}$ could be estimated based on the
maximum time steps necessary for the transmission of the electromagnetic
wave from one corner to another.

\section{Conclusion\label{sec:Conclusion}}

This paper purposes JefiPIC, a powerful plasma simulation package
that uses a 3-D particle-in-cell method in conjunction with Jefimenko's
equations to accurately model plasma systems. Despite its complexity,
we have successfully implemented JefiPIC on GPU for practical calculations.
JefiPIC is more user-friendly and easier to get started with, as it
doesn't require many complicated amendment algorithms. This makes
it an alternative option for newcomers who want to dive into the field
of particle-in-cell simulations.

Our comparative study between UNIPIC, EPOCH, and RGB-Maxwell has revealed
that JefiPIC has certain advantages in the following aspects. First,
JefiPIC's use of integral operation enables accurately employing the
linear particle-to-grid apportionment technique, which simplifies
the numerical scheme and brings less noise. Secondly, JefiPIC naturally
cuts off charge and field at the computational boundary, preventing
field reflection or charge deposition. Third, JefiPIC does not require
extra pre-processing, making it a superior choice for calculating
non-neutral plasma cases. Fourth, the integral equations used in JefiPIC
free the simulation from the CFL condition, enabling larger time steps
and helping to conserve computational resources. Finally, the results
obtained from JefiPIC are comparable to those obtained using second-order
accurate Boltzmann equations. Overall, our study demonstrates that
JefiPIC is a promising and time-efficient method for simulating plasma
physics problems with open boundary conditions.

However, to increase JefiPIC's versatility, the next version will
incorporate interactions between electrons and neutral molecules or
charged ions. Also, measures will be taken to improve the efficiency
of the integral operation or to accomplish the parallel computing.
We believe that these improvements will extend the applicability of
JefiPIC further and help advance our understanding of plasma behavior.

%%%% Acknowledgments %%%%%%%%
\section*{Acknowledgments}
We express our gratitude to Prof. Jian-Guo Wang, Associate Researcher
Zai-Gao Chen, and Assistant Researcher Ze-Ping Ren from NINT for their
valuable contributions and insightful discussions on the PIC simulations.
The work is supported by the National Key Research and Development
Program of China under Grant No. 2020YFA0709800 and the National Natural
Science Foundation of China (NSFC) under Grants No. 12105227.

\appendix

\section{The unit converion table}

In Flexible Unit (FU), all physical quantities have the dimension
of energy E, shown in Table \ref{tab:The-transform-of}. One needs
to choose proper values for $\lambda$, $c$, $\hbar$, $\varepsilon_{0}$
so that all numerical quantities are in the range of float64 on GPU,
where $c$, $\hbar$, $\varepsilon_{0}$ denote the speed of light,
the reduced Planck constant, and the vacuum permittivity. $\lambda$
is constant relating the energy quantity between SI and FU.

\begin{table}

\caption{\label{tab:The-transform-of}The transform of the physical quantities
from SI Unit to Flexible Unit}

\begin{centering}
\begin{tabular}{|c|c|c|}
\hline 
Physical quantity & SI Unit & Flexible Unit\tabularnewline
\hline 
\hline 
Magnetic field & T & $5.01398\times10^{-36}/(\lambda^{2}\hbar^{3/2}\varepsilon_{0}^{1/2}c^{5/2})\mathrm{E}^{2}$\tabularnewline
\hline 
Length & m & $3.16304\times10^{25}\hbar c\lambda\mathrm{E}^{-1}$\tabularnewline
\hline 
Time & s & $9.48253\times10^{33}\hbar\lambda\mathrm{E}^{-1}$\tabularnewline
\hline 
Electric charge & C & $1.89032\times10^{18}(\hbar\varepsilon_{0}c)1/2$\tabularnewline
\hline 
Momentum & $\mathrm{kg\cdot m\cdot s^{-1}}$ & $2.99792\times10^{8}/(c\lambda)E$\tabularnewline
\hline 
Energy & J & $1/\lambda E$\tabularnewline
\hline 
Mass & kg & $8.89752\times10^{8}/(c^{2}\lambda)E$\tabularnewline
\hline 
Electric current & A & $1.99347\times10^{-16}(\varepsilon_{0}c/\hbar)^{1/2}/\lambda E$\tabularnewline
\hline 
Electric field & $\mathrm{V\cdot m^{-1}}$ & $1.67249\times10^{-44}/(\lambda^{2}\hbar^{3/2}\varepsilon_{0}^{1/2}c^{3/2})\mathrm{E}^{2}$\tabularnewline
\hline 
Force & $\mathrm{kg\cdot m\cdot s^{-2}}$ & $3.16153\times10^{-26}/(\hbar\varepsilon_{0}c)\mathrm{E}^{2}$\tabularnewline
\hline 
Unit charge & 1 & $0.30286(\hbar\varepsilon_{0}c)^{1/2}$\tabularnewline
\hline 
\end{tabular}
\par\end{centering}
\end{table}

%%%% Bibliography  %%%%%%%%%%

\end{document}